\documentclass[aps,superscriptaddress,nofootinbib,showpacs]{revtex4}

\usepackage{amssymb}
\usepackage{amsmath}
\usepackage{amsthm}
\usepackage{slashed}
\usepackage{upgreek}

\newtheorem{thm}{Theorem}%[section]
\newtheorem{lem}[thm]{Lemma}%[section]
\newtheorem{Def}{Definition}%[section]
\newtheorem{cor}[thm]{Corollary}%[section]

\newtheorem*{thmP}{Theorem (Penrose 1965)}
\newtheorem*{thmH}{Theorem (Hawking-Penrose 1970)}

\newtheorem*{thmG}{Theorem (Gannon 1975)}
\newtheorem*{thmW}{Theorem (Witten 1994)}

\begin{document}

\title{Singularities from the Topology and Differentiable Structure of Asymptotically Flat Spacetimes}

\author{Kristin\ Schleich}
\author{Donald M.\ Witt}
\affiliation{Department of Physics and Astronomy, University of British Columbia,
Vancouver, British Columbia \ V6T 1Z1}
\affiliation{Perimeter Institute for Theoretical Physics, 31 Caroline Street North, Waterloo,
ON, N2L 2Y5, Canada}
\date{\today}

\begin{abstract}
We prove that certain asymptotically flat initial data sets with nontrivial topology and/or differentiable structure collapse to form singularities. The class of such initial data sets is characterized by a new smooth invariant, the maximal Yamabe invariant, defined through smooth compactification of the asymptotically flat manifold. Our singularity theorem applies to spacetimes  admitting a Cauchy surface of nonpositive maximal Yamabe invariant with initial data that satisfies the dominant energy condition. This class of spacetimes includes simply connected spacetimes with a single asymptotic region, a class not covered by prior singularity theorems for topological structures. The maximal Yamabe invariant can be related to other invariants including, in $4$ dimensions, the  $\widehat A$-genus and the Seiberg-Witten invariants. In particular, 5-dimensional spacetimes with  asymptotically flat Cauchy surfaces with non-trivial Seiberg-Witten invariants are singular. This singularity is due to the differentiable structure of the manifold. 
\end{abstract}
\maketitle

\section{Introduction}

In 1965, Penrose proved the first singularity theorem; under certain physically reasonable conditions, a spacetime must have an inextendible null geodesic,  that is it must be singular \cite{Penrose:1964wq}.  Precisely
\begin{thmP} Spacetime ${\cal M}$ with metric $\gamma_{ab}$ cannot be null geodesically complete if
1) The null convergence condition, $R_{ab}W^aW^b \geq 0$ for all null vectors $W^a$, holds;
2) there is a non-compact Cauchy surface $\Sigma$ in  ${\cal M}$;
3)  there is a closed trapped surface ${\cal T}$ in ${\cal M}$.
\end{thmP}
\noindent 
This profound result demonstrated that singularities  exhibited by known exact solutions such as Schwarzschild spacetime were not a consequence of their high symmetry  but rather a general feature of gravitational collapse. Generalizations of this theorem demonstrated the existence of inextendible timelike geodesics under suitable energy conditions and extended its application to a variety of other physical situations  \cite{Hawking:1966vg,Hawking:1969sw,HawkingEllis}.
In general, singularity theorems require the existence of a trapped surface or an equivalent condition that indicates the initiation of  gravitational collapse. Hence spacetimes without these structures, such as Minkowski spacetime and static star solutions, are nonsingular. In particular, spacetimes with Cauchy surfaces of ${\mathbb R}^3$ topology can be either singular or nonsingular depending on whether or not a trapped surface is present. 

This is not the case if the Cauchy surface has nontrivial topology and certain asymptotic behavior. Gannon showed that  any physically reasonable  asymptotically flat spacetime  with a non-simply connected $3$-dimensional Cauchy surface must be singular \cite{Gannon:1975}, namely
\begin{thmG} Let ${\cal M}^{4}$ be a spacetime which satisfies the null convergence condition and admits a Cauchy surface $\Sigma^3$ which is regular near infinity. If $\Sigma^3$ is non-simply connected, then ${\cal M}^{4}$ is not null geodesically complete.\end{thmG}
\noindent Subsequent generalizations of this result extended its conclusions to a broader class of physical situations in $4$ dimensions \cite{Gannon:1976,Lee:1976,Galloway:1983xx}. Furthermore, the topological censorship theorem of  Friedman, Schleich and Witt proved that the topology of physically reasonable, asymptotically flat spacetimes could not be actively probed by distant observers \cite{Friedman:1993ty}; all non-simply connected topological structures are behind horizons. These results were extended to the locally asymptotically anti-de Sitter case in \cite{Galloway:1999bp}. As the topology of $3$-manifolds are characterized by their fundamental group,\footnote{The recent proof of the Poincar\'e conjecture by Perelman removes the possibility of a homotopy 3-sphere noted in some older papers on topological censorship \cite{Perelman:2003uq,Perelman:2006up,Perelman:2006un}.} these results apply to all isolated topological structures in asymptotically flat $4$-dimensional spacetime. In fact, topological censorship completely characterizes the topology of asymptotically flat $4$-dimensional spacetime exterior to the horizons; this region is simply connected \cite{Galloway:1995xx}.

Although  the singularity theorems were initially proven for $4$-dimensional spacetimes, their results immediately generalize to  higher dimensions; Gannon's singularity theorem  can be generalized to higher dimensional spacetimes with non-simply connected, asymptotically flat Cauchy surfaces. The topological censorship theorems also hold in higher dimensions \cite{Galloway:1999br}. However, in $5$ or more spacetime dimensions, the topology of the Cauchy surface is no longer completely characterized by its fundamental group. For example,  all  simply connected $4$-manifolds  are connected sums of $S^4$, $S^2\times S^2$, ${\mathbb C}P^2$, $\overline{{\mathbb C}P^2}$ and $E8$ factors.  Puncturing any such smooth manifold results in a noncompact smooth 4-manifold.\footnote{Note that
$E8$ does not admit a differentiable structure; however certain connected sums containing it do. In particular $K3$, a smooth manifold, is homeomorphic to the connected sum of two $E_8$ and three $S^2\times S^2$ factors. }  This manifold can be taken to be the Cauchy surface of some globally hyperbolic $5$-dimensional spacetime as  it admits  asymptotically flat initial data satisfying the dominant energy condition \cite{Witt:1986ng,Witt:2009za}. Consequently, there are $5$-dimensional spacetimes with nontrivial topological structures  that evade the conditions of Gannon's theorem and the topological censorship theorem.  This is also true in $6$ or more spacetime dimensions. Therefore these theorems leave open the issue of whether or not all topological structures collapse to form singularities in $5$ or more dimensions.

 This paper addresses this issue; we show that a certain class of  topological structures in $5$ or more spacetime dimensions collapse to form singularities. Specifically, we prove a new singularity theorem,  Theorem \ref{jangsing}, for spacetimes with Cauchy surfaces of topology and/or differentiable structure in a specified class with  asymptotically flat initial data that satisfies the dominant energy condition.  This class is defined through
 a natural extension of the Yamabe invariant for compact manifolds to the asymptotically flat case. 
An asymptotically flat $n$-manifold is related to a closed $n$-manifold 
by attaching $n$-balls to each asymptotic region via smooth attaching maps. The Yamabe invariant  of  the asymptotically flat manifold  is defined to be that of the resulting closed manifold. This definition, in general dimension, depends on the choice of attaching maps. To remove this dependence, the maximal Yamabe invariant is defined as the supremum over all possible attaching maps. Theorem \ref{jangsing} applies to this class of spacetimes, that is ones whose Cauchy surfaces have nonpositive maximal Yamabe invariant. Included in this class are simply connected  Cauchy surfaces  with nontrivial topology in $5$ or more spacetime dimensions. Consequently, Theorem \ref{jangsing} applies to a class of spacetimes not addressed by the generalization of Gannon's theorem to higher dimensions.
 
  Our approach to proving the singularity theorem  is to demonstrate that the Cauchy surface  must exhibit one or more apparent horizons. To do so involves two key results. We first prove Theorem \ref{compactify2}: an asymptotically flat n-manifold with nonnegative scalar curvature has positive maximal Yamabe invariant. Next,
we prove Theorem \ref{jangpos}: if  an asymptotically flat initial data set satisfying the dominant energy condition has a global solution to the Jang equation, then the Cauchy surface admits an asymptotically flat metric with zero scalar curvature.

  The singularity theorem, Theorem \ref{jangsing} then follows from these two theorems and the existence of solutions to the Jang equation  \cite{Jang:1978}. 
Shoen and Yau  proved the existence of solutions to the Jang equation in  \cite{Schon:1981vd} as part of their proof of the positive energy theorem. Furthermore,  obstructions  to a global solution imply that the initial data set contains apparent horizons. Although \cite{Schon:1981vd} explicitly treats only the case of $3$ dimensional Cauchy surfaces, these results can be extended through $7$ dimensions using  \cite{schoen1975} and \cite{Eichmair:2007}. 
Theorem \ref{jangsing} follows by contradiction: Assume that there is a global solution to the Jang equation on the Cauchy surface with nonpositive maximal Yamabe invariant.  Theorem \ref{jangpos} then implies that the Cauchy surface admits an asymptotically flat metric of zero scalar curvature. But this implies that the maximal Yamabe invariant is positive by Theorem \ref{compactify2}, in contradiction. It follows that there is not a global solution to the Jang equation; therefore the initial data set contained one or more apparent horizons. Hence the spacetime is singular.  

This approach is similar in spirit to that used in the generalization of Gannon's theorem by Galloway \cite{Galloway:1983xx}; however, Galloway's result uses a result of Meeks, Simon and Yau on the existence of minimal surfaces that applies only to $3$-manifolds \cite{Meeks:1982xx}. Our singularity theorem  for noncompact Cauchy surfaces applies to  any $d$-dimensional spacetime, $3\leq d\leq 8$, with nonpositive maximal Yamabe invariant that admits asymptotically flat initial data satisfying the dominant energy condition. The dominant energy condition is more restrictive than the null convergence condition; hence our theorem applies to a more restrictive set of spacetimes than Gannon's singularity theorem  and the topological censorship theorem. However the class of  structures covered by Theorem \ref{jangsing} contains a set of simply connected Cauchy surfaces with a single asymptotic region - topologies that other topological singularity theorems do not address.  Consequently, our theorem establishes that there is a class of simply connected spacetimes in $5$ or more dimensions that collapse to form singularities. In particular, there are an infinite number of $4$-dimensional simply connected asymptotically flat Cauchy surfaces with nonpositive maximal Yamabe invariant. In $4$ dimensions, a nonpositive maximal Yamabe invariant is related to nonvanishing $\widehat A$-genus and to nonvanishing Seiberg-Witten invariants. As the Seiberg-Witten invariants can be used to characterize exotic differentiable structures on non-spin $4$-manifolds,  $5$-dimensional simply connected spacetimes can collapse due to either their topology or their differentiable structure.
 
The roadmap of the paper is as follows: Section \ref{prelim} provides a summary of basic definitions and theorems. Section \ref{yamabe} defines the maximal Yamabe invariant for asymptotically flat $n$-manifolds. The required result on the scalar curvature of the compactified manifold, Theorem \ref{compactify2}, is proven in Section \ref{sect4}. Theorem \ref{jangpos} and the singularity theorem are proven in Section \ref{sect5-jang}. Section \ref{sect6} relates nonpositive maximal Yamabe invariant  to the $\widehat A$-genus and to the Seiberg-Witten invariants and gives families of examples of such smooth, simply connected asymptotically flat $4$-manifolds. We conclude with a discussion in Section \ref{sect7}.

\acknowledgments  The authors would like to thank KITP for its hospitality in 
February, 2003 where an early version these results, available at http://online.itp.ucsb.edu/online/joint98/schleich/, was first presented.  Part of this work was also presented at the Black Holes IV: Theory and Mathematical Aspects, 
Honey Harbour, Ontario, May 2003 and the Canadian Mathematical Society Summer meeting, Edmonton, Alberta, June 2003.
The authors would also like to thank the Natural Sciences \& Engineering Research 
Council of Canada (NSERC) for financial support and Perimeter Institute where this work was
expanded. 

 \section{Preliminaries}\label{prelim}

We begin by giving some basic definitions needed in the statement and proof of the theorems in subsequent sections.
An {\it initial data set} for the Cauchy problem in general relativity consists
of a $n$-manifold $\Sigma ^n$ that is geodesically complete with respect to riemannian metric $g_{ab}$, a  symmetric tensor
$p_{ab}$, energy density $\mu $, and momentum
density $J^a$. These fields satisfy the Hamiltonian and momentum constraints
\begin{align}R-p_{ab}p^{ab}+p^2&=2\mu  \, \label{hconstraint}\\
D_b(p^{ab}-pg^{ab})&={J^a} \  \label{mconstraint} \end{align}
where $R$ is the scalar curvature of the metric $g_{ab}$, $D_b$ is the covariant
derivative defined with respect to $g_{ab}$, and $p=g^{ab}p_{ab}$.
In addition,  the fields are required to be sufficiently regular; for convenience, they will assumed to be  smooth ($C^\infty$), though the results of this paper readily generalize to sufficiently differentiable fields.

{\it Physically reasonable}  initial data obeys a 
local energy condition, the {\it dominant energy condition} (DEC),
namely,  $\mu \ge {\sqrt {(J_aJ^a)}}$.
When the energy and momentum densities correspond to vacuum or classical, nondissipative matter
sources,  local existence theorems show that the initial data evolves under the Einstein or coupled Einstein-matter equations
into a globally hyperbolic spacetime with topology ${\mathbb R}\times \Sigma^n$ \cite{HawkingEllis2}.
In this spacetime, $\Sigma^n$ is a  {\it Cauchy surface}, a spacelike hypersurface 
such that every non-spacelike curve intersects it exactly once; $p_{ab}$ and $g_{ab}$ are now identified with the extrinsic curvature  and  induced metric of this Cauchy surface. From this point on, initial data sets will be assumed to be physically reasonable initial data sets.

The topology of the manifold $\Sigma^n$ is not restricted in the definition of an initial data set. However, there are two cases of particular interest; initial data sets on closed manifolds,\footnote{A $n$-manifold is {\it closed} if it is compact and 
has no boundary. } describing cosmological models, and those on asymptotically flat manifolds, describing isolated gravitational systems.
The asymptotically flat case is the focus of this paper.

Precisely, $\Sigma^n$ is an {\it asymptotically flat} $n$-manifold if,  for some
compact smooth submanifold with boundary $N^n\subset \Sigma ^n$, $\Sigma ^n - N^n$ 
consists of a finite number of  disconnected components, each of which is diffeomorphic to ${\mathbb R}^n$
minus a n-ball, ${\mathbb R}^n -B^n$.\footnote{
 $B^n=\{ x\in {\mathbb R}^n|\  ||x|| \leq 1\}$.} Furthermore, ${\partial N}^n$ is a finite disjoint
union of smooth $(n-1)$-spheres, ${\partial N}^n=\coprod_k S^{n-1}$.

This definition of an asymptotically flat $n$-manifold is given only in terms of the properties of differentiable manifolds and does not require any
additional structure. It does not, in itself, restrict the metric, connection, or any other geometric structure on $\Sigma^n$ in any way. 
All asymptotically flat $n$-manifolds are, up to diffeomorphism, closed, smooth $n$-manifolds with points removed: every $\Sigma^n$ can be obtained from a closed smooth
$n$-manifold ${\tilde {\Sigma }^n}$ via removing a finite number of points.  Each removed point has a neighborhood diffeomorphic to ${\mathbb R}^n -B^n$; these neighborhoods are asymptotic regions.  Conversely, ${\Sigma ^n}$ can be smoothly 
compactified to a smooth manifold ${\tilde {\Sigma}} ^n$
by adding a finite set of isolated points  to compactify each asymptotic region.
For fewer than three dimensions, all topological manifolds are smoothable
and any two which are homeomorphic are diffeomorphic. However, this is no longer the case  in four or more dimensions; there exist topological $n$-manifolds that are not smooth; i.e. they do not admit a differentiable structure. Furthermore, topological $n$-manifolds that admit a differentiable structure may admit additional differentiable structures that are not diffeomorphic to each other.  Thus, in four or more dimensions, the process of compactification of an asymptotically flat manifold requires careful specification. The details
of this will be given in section \ref{yamabe}.

A metric is an {\it asymptotically flat metric} if the pullback of $g_{ab}$, ${\hat g}_{ab}$, from $\Sigma ^n- N^n$ onto each neighborhood ${\mathbb R}^n - B^n$ satisfies ${\hat g}_{ab}- \delta_{ab}= O(\frac{1}{ r^{n-2}})$, 
$\partial_c{\hat g}_{ab}=O(\frac{1}{ {r^{n-1}}})$, 
$\partial_d \partial_c {\hat g}_{ab}=O(\frac{1}{ {r^{n}}})$ as $r\to \infty$ where $\delta_{ab}$ is the flat metric.
An initial data set  is  an {\it asymptotically
flat initial data set} if $\Sigma^n$ is an asymptotically flat $n$-manifold with asymptotically flat metric $g_{ab}$ and the pullback of $p_{ab}$,  ${\hat p}_{ab}$, from  $\Sigma^n$ onto each ${\mathbb R}^n - B^n$ satisfies
${\hat p}_{ab}=O(\frac{1}{{r^{n-1}}})$, 
 $\partial_c {\hat p}_{ab}=O(\frac{1}{ {r^{n}}})$ as $r\to \infty$. The pullbacks of the energy and momentum densities,
$\hat \mu$ and $\hat J^a$ respectively, satisfy  fall-off conditions as required to satisfy the constraints.  The convention adopted in this paper is that 
initial data on asymptotically flat $n$-manifolds is asymptotically flat  initial data unless stated otherwise.

As first shown in \cite{Witt:1986ng}, one can construct asymptotically flat initial data sets  which
obey the dominant energy condition on any asymptotically flat manifold $\Sigma^n$.\footnote{See also the generalization to the vacuum case in \cite{Chrusciel:2004cc}.} Therefore, the Einstein equations place
no restriction on the choice of the topology of $\Sigma ^n$. 

We next outline the proof of a  theorem, needed for later results, that demonstrates the existence of conformally related metrics with zero scalar curvature on asymptotically flat manifolds. Various forms of this result have been proven in three dimensions by several authors elsewhere \cite{Schon:1979rg, Cantor:1981, Witt:2009za}. The treatment below, generalized to $n$ dimensions,  is given in detail in three dimensions in \cite{Witt:2009za}. 

First, the conformal Laplacian operator $L$ in $n$ dimensions is given by
\begin{align}
L &= -a_n{D}^2 + { R} \label{conflaplacian}\\
a_n&=\frac{4(n-1)}{n-2} 
\end{align}
where $D^2=D_aD^a$, and $D_a$ and $R$ are, respectively, the covariant derivative and scalar curvature of the metric $g_{ab}$.

\begin{lem}\label{CLsolution} Let $\Sigma^n$ be an asymptotically flat manifold with smooth asymptotically flat metric $g_{ab}$. If $L$ is positive on smooth functions with compact support,  then there is a smooth positive solution of $L\phi=0$ such
that $\phi\to 1$ with asymptotic fall-off as $r\to\infty$ in every asymptotic region of $\Sigma^n$.
\end{lem}

\begin{proof}

On smooth functions with compact support, $\phi,\psi \in C^{\infty }_0({\Sigma }^n)$, define
\begin{equation*} %\label{innerprod}
(\psi, \phi )_L=\int _{{\Sigma }^n} d{\mu }_{ g}(a_n{D}_a\psi {D}^a\phi + {R}{\psi }{\phi })
\end{equation*}
As $L$ is positive for  $\phi \in C^{\infty }_0({\Sigma }^n)$, it follows that $(\phi, \phi )_L>0$ and 
$(\phi, \phi )_L=0$ if and only if $\phi=0$. Hence, this is an inner product  on $C^{\infty }_0({\Sigma }^n)$.  The completion of this inner product yields a Hilbert space ${\cal H}_L$.

Consider the equation
\begin{equation}\label{flatconf}
-a_n{ D}^2\psi + {\bar R}{\psi }= -{\bar R}
\end{equation}
on ${\cal H}_L$. 
Define the functional $F: {\cal H}_L\rightarrow \mathbb{R}$ 
\begin{equation*}
F(\phi )= -\int _{\Sigma ^n} d{\mu }_{g} {R}\phi \ . \end{equation*}

Note that there is a constant $K>0$ such that $|| \phi|| < K||\phi||_L$ where
\begin{equation*}
 || \phi|| = \Bigl( \int _{\Sigma ^n} d\mu_{g}( D_a\phi D^a \phi + \phi^2)\Bigr)^{\frac{1}{2}}
\end{equation*}
is the usual norm on ${\cal H}_1$
and $||\phi||^2_L=(\phi,\phi)_L$. 
Consequently  $||\phi||_2= (\int _{\Sigma ^n} d\mu_{g} \phi^2)^{\frac 12} \leq  || \phi|| \leq K||\phi||_L$. This implies that
$F(\phi)$ is a bounded functional,
\begin{equation*}
|F(\phi)|\leq  \int _{\Sigma ^n} d{\mu }_{g} |{R}\phi|\leq ||R||_2||\phi||_2 \leq C||\phi||_L \ .
\end{equation*}
using Holder's inequality, 
$ \int_{\Sigma^n} d\mu_{g} |{R}{\phi }|\leq ||R||_q||\phi||_{ p} $,
 with $p=q=2$ and the relationships of the norms $||\phi||$ and $||\phi||_L$.
Therefore, the Riesz representation theorem (See, for example \cite{Gilbarg}) implies that there is a unique $\psi \in {\cal H}_L$ such that
\begin{equation*}
(\phi, \psi)_L=F(\phi )
\end{equation*}
for all $\phi \in {\cal H}_L$. In particular, this is true for all $\phi \in C^{\infty }_0({\Sigma }^n)$; consequently 
$\psi $ is a weak solution to (\ref{flatconf}).
Moreover, the solution is smooth by regularity 
theorems for second order elliptic operators (See, for example \cite{Gilbarg}). In addition, the pullback of $\psi$ to ${\mathbb R}^n - B^n$  vanishes with fall-off $\hat \psi =O(\frac{1}{ r^{n-2}})$
as $r\to \infty$ in each asymptotic region because of the asymptotic behavior of $g_{ab}$ and consequently that of $L$ and $R$. Consequently, the function $\phi = 1+\psi $ solves 
\begin{equation*}%\label{solutionzero}
-a_n{\bar D}^2\phi + {\bar R}{\phi }= 0\ 
\end{equation*}
and $\phi\to 1$ with asymptotic fall-off of its pullback $\hat \phi=1+O(\frac 1{r^{n-2}})$ as $r\to\infty$ in each asymptotic region. Furthermore one can show $\phi >0$ by smoothness and  application of the maximum principle  \cite{Spivak}. 
\end{proof}

\begin{thm}\label{conftoR=0} Let $\Sigma^n$ be an asymptotically flat manifold with smooth asymptotically flat metric $g_{ab}$. If $L$ is positive on smooth functions with compact support, then there is an asymptotically flat metric  $g'_{ab}$ conformally related to $g_{ab}$ on $\Sigma^n$ with vanishing scalar curvature.
\end{thm}
\begin{proof}
The conformally related metric $g'_{ab}={\phi }^{\frac 4{n-2}}{ g}_{ab}$ has scalar curvature
\begin{equation*}
R'={\phi }^{- \frac{n+2}{n-2}}\bigl(-a_n{ D}^2\phi  
+ {R}\phi \bigr) \\
\end{equation*}
where $D^2=D_aD^a$, and $D_a$ and $R$ are respectively the covariant derivative and scalar curvature of $g_{ab}$.
This curvature vanishes, $R'=0$, if  there is a smooth positive  solution  of $L\phi=0$ where $L$ is the conformally invariant laplacian operator (\ref{conflaplacian}). As $L$ is positive, such a solution exists by Lemma \ref{CLsolution}. 
Furthermore, as  $\phi >0$ and has asymptotic fall-off $\hat \phi=1+O(\frac 1{r^{n-2}})$ as $r\to\infty$, the conformally related metric $g'_{ab}$ is asymptotically flat.
\end{proof}

\section{A Yamabe invariant for asymptotically flat manifolds} \label{yamabe}

A well known characterization of the allowed scalar curvature of geometries on a closed $n$-manifold is given by the Yamabe invariant:
\begin{Def}\label{yambinvar}
The Yamabe invariant ${\sigma }(M^n)$ for a closed $n$-manifold $M^n$, $n\geq 2$, is  
\begin{equation}\label{cyamabe}
{\sigma }(M^n)= \sup _{g\in {\rm Riem}(M^n)} {\cal Y}(g)
\end{equation} 
 where $\displaystyle
{\cal Y}(g)= \inf _{f\in C^{\infty}(M^n)} {\cal E}(e^{2f}g)$, $\displaystyle
 {\cal E}(g)=\frac { \int _{M^n} R_g d{\mu _g}}{\bigl( \int _{M^n}  d{\mu _g}\bigr)^{\frac{n-2}{n}}}$,
and ${\rm Riem} (M^n)$ is the space of smooth riemannian metrics on $M^n$.
\end{Def}

For $2$-manifolds, the Yamabe invariant is  simply proportional to the Euler characteristic, ${\sigma }(N^2)= 4{\pi }{\chi }(M^2)$.  All closed $2$-manifolds admit a metric of constant curvature by the uniformization theorem; thus the Yamabe invariant, or equivalently the Euler characteristic, fixes the sign of the curvature. Therefore, a closed $2$-manifold admitting a metric of
constant curvature of one sign can not admit one of a different sign; there is a topological obstruction.

 This is no longer the case in three or more dimensions. Again  any closed 
manifold $M^n$, $n\geq 3$, admits a metric or metrics with constant scalar curvature. However, if $M^n$ admits a metric with positive
scalar curvature, then it also admits metrics of constant scalar curvature of all signs. Consequently, if   ${\sigma }(M^n)>0$, there is no obstruction to a metric of constant scalar curvature of any sign. If  $M^n$ admits a metric with zero constant scalar curvature 
but not one with positive constant scalar curvature, then ${\sigma }(M^n)=0$. Consequently, $M^n$ will admit metrics with zero or negative constant scalar curvature, but not ones with positive constant scalar curvature. 
Finally, if $M^n$ admits only metrics with negative constant scalar curvature, then ${\sigma }(M^n)\leq 0$.\footnote{Zero Yamabe invariant may occur as the supremum need not be attained in the class of metrics.}
We now generalize the Yamabe invariant to asymptotically flat manifolds. The process is to first construct a closed $n$-manifold by a smooth compactification of the asymptotically flat $n$-manifold. The Yamabe invariant of this closed manifold is simply (\ref{cyamabe}). The generalized Yamabe invariant for asymptotically flat manifolds is then defined in terms of  the supremum of the Yamabe invariant of the closed manifolds obtained by all possible smooth compactifications of the asymptotically flat manifold.

Before proceeding, it is useful to  motivate this definition with a simple example. It is well known that  the one point compactification of ${\mathbb R}^n$ is homeomorphic to $S^n$ for any $n$.  It is also well known that for $n\geq 7$, $n$-spheres that are homeomorphic are not necessarily diffeomorphic; the $7$-sphere has $28$ inequivalent differentiable structures, the $8$-sphere has $2$ and the $9$-sphere has $8$.  Furthermore, there are obstructions to positive scalar curvature on certain exotic  $9$-spheres and $10$-spheres \cite{Adams, Milnor,Hitchin}. Therefore, the one point compactification of a manifold does not always provide sufficient information about the possible differentiable structures on the compactified manifold needed for the Yamabe invariant even in this simple case.  In other words, the Yamabe invariant for a closed manifold is not invariant under homeomorphisms of the manifold in all dimensions; it depends on the differentiable structure. 
 Additionally, restricting the compactification in some way so that it yields a unique differentiable structure may itself restrict the Yamabe invariant  in some unknown way.
Therefore, a robust generalization of the Yamabe invariant to asymptotically flat manifolds using compactification should recognize that there is more than one way to smoothly compactify each asymptotic region. The approach taken in this work is to compactify the asymptotically flat $n$-manifold by attaching not simply a point, but  a neighborhood  of the point and concretely specifying the attachment of this neighborhood to the asymptotic region. This process precisely characterizes possible smooth structures on the compactification.

To begin, recall the standard definition of attaching two manifolds with boundary in which one
smoothly glues the manifolds together along their boundaries with the gluing given
by a specified diffeomorphism \cite{Hirsch}:

\begin{Def}\label{attach}
Given two smooth $n$-manifolds $P^n$ and $Q^n$ with boundaries $\partial P^n$ and $\partial Q^n$ and
a diffeomorphism  $f:\partial P^n\rightarrow \partial Q^n$, the {\rm smooth adjunction space} ${W^n}_{f }$
is 
\begin{equation*}
{W^n}_{f }=P^n\cup _{f } Q^n\equiv {\frac {P^n\coprod Q^n}\sim }
\end{equation*}
where ${P^n\coprod Q^n}$ is the disjoint union of $P^n$ and $Q^n$ and the equivalence relation $\sim $ in the identification 
is given by $x\sim f(x)$ $\forall x\in \partial Q^n$. 
\end{Def}

Two basic properties of the smooth adjuction space and its dependence on the diffeomorphism $f$
are given by the following theorem and its corollary \cite{Hirsch}:

\begin{thm}\label{deformequiv} Let $f_0:\partial Q_0^n\rightarrow \partial P^n$ and 
$f_1:\partial Q_1^n\rightarrow \partial P^n$ be diffeomorphisms. Suppose that the diffeomorphism
$f_1^{-1}f_0:\partial Q_0^n\rightarrow \partial  Q_1^n$ extends to a diffeomorphism 
$h: Q_0^n\rightarrow  Q_1^n$. Then the two $n$-manifolds $P^n\cup _{f_0} Q_0^n$ and $P^n\cup _{f_1 } Q_1^n$ 
are diffeomorphic.
\end{thm}

\begin{cor}\label{equiv} Let $P^n$ and $Q^n$ be two smooth $n$-manifolds with respective boundaries 
$\partial P^n$ and $\partial Q^n$. If  $f:\partial P^n\rightarrow \partial Q^n$ and 
$g:\partial P^n\rightarrow \partial Q^n$ are isotopic diffeomorphisms, then the two $n$-manifolds
$P^n\cup _{f } Q^n$ and $P^n\cup _{g } Q^n$ are diffeomorphic.
\end{cor}

Given the above, we can define the smooth compactification
of an asymptotically flat manifold:

\begin{Def}\label{asymcomp}
The smooth compactification ${\tilde {\Sigma }^n}_{\Phi }$
of an asymptotically flat manifold $\Sigma ^n$
\begin{equation*}
{\tilde {\Sigma }^n}_{\Phi }=N^n\cup _{\Phi } {\cal I}_0\equiv {\frac {N^n\coprod  {\cal I}_0}\sim }
\end{equation*}
where ${\cal I}_0$ is a disjoint union of n-balls $B^{n}_i$'s and $\Phi $ is a finite set 
of diffeomorphisms $\{ \phi _i \}$ where $\phi _i: S^{n-1}_i=\partial B^{n}_i\rightarrow \partial N^n$, indexed
by $i$. $N^n\coprod {\cal I}_0$ is the disjoint union of $N^n$ and ${\cal I}_0$ and the equivalence relation for identification 
is given by $x\sim \phi _i(x)$, $\forall x\in S^{n-1}_i$. 
\end{Def}

Corollary \ref{equiv} implies that ${\tilde {\Sigma }^n}_{\Phi }$ is determined,
up to diffeomorphism, by the isotopy classes of diffeomorphisms $\phi_i\in \Phi$.  It immediately follows that as the boundary $\partial N^n$ is just  the disjoint union of spheres $S^{n-1}$,
the isotopy classes  are given by $\pi _ 0Diff(S^{n-1})=Diff(S^{n-1})/Diff_{id}(S^{n-1})$ for the smooth compactification of an asymptotically flat $n$-manifold. 

Representing the sphere $S^{n-1}$ by its embedding in Euclidean space, $S^{n-1}=\{(x^1,x^2,\ldots,x^n)\in { \mathbb R}^n| \sum_{i=1}^n (x^i )^2=1\}$,  the map $P:(x^1,x^2,\ldots,x^n)\to(-x^1,x^2,\ldots,x^n)$ for points on $S^{n-1}$ is an  orientation reversing  $O(n)$ isometry. This isometry trivially extends to the interior of  the n-ball $B^n$. Its composition with any diffeomorphism  in the identity component of the diffeomorphism group, also extends trivially.  Consequently,  by Theorem \ref{deformequiv}, the smooth adjuction space formed by attaching an n-ball with $f$ is diffeomorphic to that formed by doing so with $Pf'$ for any  $f,f'\in Diff_{id}(S^{n-1})$. Hence two smooth adjunction spaces formed by attaching with diffeomorphisms $g$  and $g'$ respectively such that $g^{-1}g'=f$ or  $g^{-1}g'=Pf$ are diffeomorphic. In other words,  attaching with diffeomorphisms in two isotopy classes  equivalent under the orientation reversing map result in diffeomorphic adjuction spaces. Therefore, it suffices to consider  $\phi_i\in \pi _ 0Diff^+(S^{n-1})$, the isotopy classes of orientation preserving diffeomorphisms.

\begin{Def}\label{asymyambconst} Let $\Sigma ^n$ be an asymptotically flat $n$-manifold. The asymptotically flat
Yamabe invariant, ${\hat {\sigma }}(\Sigma ^n,\Phi )$, is defined by ${\hat {\sigma }}(\Sigma ^n,\Phi )=
{\sigma }({\tilde \Sigma }_\Phi ^n)$ where ${\tilde \Sigma }_\Phi ^n$ is the smooth compactification
of the asymptotically flat manifold $\Sigma ^n$ obtained using the finite set of attaching maps $\Phi $ 
and ${\sigma }({\tilde \Sigma }^n_\Phi)$ is the usual Yamabe invariant for the resulting closed $n$-manifold.
\end{Def}

It is  useful to define an additional constant
independent of the attaching map:

\begin{Def}\label{asymyamb} Let $\Sigma ^n$ be an asymptotically flat $n$-manifold. The asymptotically flat
maximal Yamabe invariant, ${\hat {\sigma }}(\Sigma ^n)$, is defined by 
\begin{equation*}
{\hat {\sigma }}(\Sigma ^n)=\sup _{\Phi \in \pi _0Diff^+(S^{n-1}) } {\hat \sigma }({ \Sigma }^n,\Phi ). 
\end{equation*} 
\end{Def}
Note that the supremum is taken over all possible choices of the isotopy class of each $\phi_i$ in  $\Phi$. 
Whether or not ${\hat {\sigma }}(\Sigma ^n,\Phi )$ actually depends on $\Phi $ changes with dimension.  For  $n=1,2,3,4$ and $6$, there is only one isotopy class,  $\pi _ 0Diff^+(S^{n-1})=1$, and smooth compactification results in a  $\tilde \Sigma^n$ that is unique up to diffeomorphism. Thus ${\hat {\sigma }}(\Sigma ^n,\Phi )$ is unique for $n=1,2,3,4$ and $6$. However, for $n\geq 7$ the isotopy classes are not trivial: $\pi _ 0Diff^+(S^6)$ has $28$ elements, $\pi _ 0Diff^+(S^7)$ has $2$ and $\pi _ 0Diff^+(S^8)$ has $8$. Thus ${\hat {\sigma }}(\Sigma ^n,\Phi )$ may now depend on the choice of the isotopy class of each $\phi_i$ in  $\Phi$. In other words, different choices of $\Phi$ may result in compactified manifolds with the same topology but inequivalent differentiable structures.\footnote{This  is  why 
$S^7$, $S^8$ and $S^9$ have their respective number of inequivalent differentiable structures.} 

The dimension of $\pi _ 0Diff^+(S^4)$ is unknown, so whether or not $\hat\sigma( \Sigma^5,\Phi)$  is unique is not determined by this construction. However, its uniqueness can be established by an alternate realization of the compactification. Observe that every homotopy $n$-sphere  with $n\geq 5$ is obtained from attaching two closed n-balls  along $\partial B^n=S^{n-1}$ via a diffeomorphism in $Diff^+(S^{n-1})$. Hence, 
\begin{equation*}
{\tilde \Sigma }^n _{\Phi }={\tilde \Sigma }^n _{id }\#F^n_1\#F^n_2\dots \#F^n_k
\end{equation*} 
where ${\tilde \Sigma }^n _{id }$ is the compactification of $\Sigma^n$  with all attaching maps in the identity of $Diff^+(S^{n-1}) $, 
$F^n_i\in \Theta _n$,  the group of homotopy $n$-spheres, and $k$ is the number of asymptotic regions in $\Sigma^n$. Now, $\Theta _n=1$
for $n=5,6$; it follows that $
{\hat \sigma }({ \Sigma }^n,\Phi )
$ is unique for $n\leq 6$. Moreover, for $n\geq 7$, this construction yields an alternate expression for $\hat \sigma(\Sigma^n)$  in terms of a connected sum with homotopy $n$-spheres.  In summary,

\begin{thm}\label{trivialunique}  Let $\Sigma ^n$ be an asymptotically flat $n$-manifold.  For $n\leq 6$, 
$
{\hat \sigma }({ \Sigma }^n,\Phi )
$ is unique and $ \hat \sigma (\Sigma^n) ={\hat \sigma }({ \Sigma }^n,\Phi ) $.
 For $n\geq 7$,
\begin{equation*}
{\hat {\sigma }}(\Sigma ^n)= \sup _{\{F^n_i\}\in \Theta _n}{\sigma }({\tilde \Sigma }^n _{id }\#F^n_1\#F^n_2\dots \#F^n_k) \label{alty}
\end{equation*} 
where ${\tilde \Sigma }^n _{id }$ is the compactification of $\Sigma^n$  with all attaching maps in the identity of $Diff^+(S^{n-1}) $,   $k$ the number of asymptotic regions and  the supremum is over all choices of $\{F^n_i\}$, a set of  $k$ elements of $\Theta _n$, the group of homotopy $n$-spheres. 

\end{thm}

Using this result, it is easy to see that the maximal Yamabe invariant is always positive for ${\mathbb R}^n$ in any dimension.  However, as discussed in detail in Section \ref{sect6} for the case of $4$ dimensions,  manifolds of more complicated topology and/or differentiable structure will have nonpositive maximal Yamabe invariant.

\section{A Compactification theorem}\label{sect4}

We now prove a compactification theorem for
asymptotically flat manifolds with asymptotically flat metrics of non-negative scalar curvature; this result is needed for the proof of the singularity theorem. For clarity of notation, we first do so for the case of one asymptotic region. 
\begin{thm}\label{compactify} Given an asymptotically flat manifold $\Sigma ^n$ with one asymptotic region and asymptotically flat metric  $g_{ab}$ with
$R\geq 0$, then ${\hat {\sigma }}(\Sigma ^n)>0$.
\end{thm}

\begin{proof}

If $R\geq 0$, then $L$, the conformal laplacian operator  (\ref{conflaplacian}), is positive on smooth functions of compact support. Theorem  \ref{conftoR=0} implies that there exists an asymptotically flat, conformally related metric $g'_{ab}={\phi }^{\frac 4{n-2}}{ g}_{ab}$ with zero scalar curvature on ${\Sigma }^n$.  For simplicity of notation, let $g_{ab}$ denote this conformally rescaled metric from this point onward. 

Let $t>0$ be a parameter chosen such that $r=t$ is a smooth $(n-1)$-sphere entirely contained in the asymptotic region ${\mathbb R}^n-B^n$. Define a new family of metrics: 
\begin{equation*}
%\label{posmetric}
g_{ab}(t)=
\begin{cases}
{ g}_{ab} & \Sigma^n\  {\rm interior}\ {\rm to}\  r< 2t\\
\alpha _t(r) \delta _{ab}+ (1-\alpha _t(r)){ g}_{ab}& 2t<r< 3t\\
\delta _{ab}& r\geq 3t\ .\\
\end{cases} 
\end{equation*}
\noindent where $\alpha _{t}$ is a family of smooth functions equal to $1$ for $r>3t$ and 0 for $r<2t$. Additionally, choose $\alpha _{t}$  to obey the fall-off conditions $|\alpha _{t}'(r)|\leq A/r^{n-2}$ 
and $|\alpha _{t}''(r)|\leq A/r^{n-1}$ where $'$ denotes the derivative with respect to $r$ and  $A$ is a constant independent of $t$. This choice can always be made.

The family of metrics $g_{ab}(t)$ is asymptotically flat and complete on ${\Sigma }^n$. For $r>3t$, the metric is, in fact, flat. The scalar curvature of $g_{ab}(t)$, $R_t$, is
only nonzero in the deformation region $2t<r<3t$. In fact, $R_t$
curvature will typically be negative in the region of deformation. Thus the conformally invariant Laplacian
operator for this family of metrics,
\begin{equation*}
L_t\phi \equiv -a_n{D_t}^2\phi + {R_t}{\phi } \ ,
\end{equation*}
is no longer manifestly positive. However, it is possible to show that it is positive for sufficiently small negative curvature in the deformation region.
This is done by appropriately choosing $t$. 
 We now show that there is a $t_0$ such that for $t>t_0$, $L_t$ is positive for $\phi\in C^\infty_0(\Sigma^n)$.
First
\begin{equation*}
\int_{\Sigma^n}d\mu_{g_t}\phi L_t \phi= \int_{\Sigma^n} d\mu_{g_t}(a_n(D_t\phi)^2 + {R_t}{\phi }^2) \  
\end{equation*}
 for $\phi\in C^\infty_0(\Sigma^n)$. Next observe that 
\begin{equation*}
  \int_{\Sigma^n} d\mu_{g_t} {R_t}{\phi }^2 =  \int_{pos} d\mu_{g_t} {R_t}{\phi }^2- \int_{neg} d\mu_{g_t} |{R_t}{\phi }^2|\geq -\int_{\Sigma^n} d\mu_{g_t} |{R_t}{\phi }^2|
\end{equation*}
where the domains of integration $pos$ and $neg$ are the support of $R_t\geq 0$ and $R<0$ in $\Sigma^n$ respectively.
It follows that
\begin{equation*}
\int_{\Sigma^n}d\mu_{g_t}\phi L_t \phi\geq \int_{\Sigma^n} d\mu_{g_t}a_n(D_t\phi)^2 -  \int_{\Sigma^n} d\mu_{g_t} |{R_t}{\phi }^2|\  .
\end{equation*}
The Sobolev inequality 
\begin{equation*}
{\Bigl( \int _{\Sigma ^n} d \mu_{g_t} |\phi |^p\Bigr)}^{\frac{2}{p}}\leq
K {\Bigl( \int _{\Sigma ^n} d\mu_{g_t} D_a\phi  D^a\phi  \Bigr)}
\end{equation*}
 where $\phi \in C^{\infty }_0({\Sigma }^n)$ and ${p=\frac{2n}{{n-2}}}$  applied to the first term on the right hand side and  Holder's inequality 
 \begin{equation*}
  \int_{\Sigma^n} d\mu_{g_t} |{R_t}{\phi }^2|\leq ||R_t||_q||\phi^2||_{\tilde p} 
  \end{equation*}
  where $\frac 1q + \frac 1{\tilde p}=1$
  applied to the second term on the right hand side  yields
  \begin{equation*}
\int_{\Sigma^n}d\mu_{g_t}\phi L_t \phi\geq\frac{a_n}{K} ||\phi||^2_p- ||R_t||_q||\phi^2||_{\tilde p} \  .
\end{equation*}
  Now
  \begin{equation*}
  ||\phi^2||_{\tilde p} =\left(\int_{\Sigma^n}d\mu_{g_t} |\phi^2|^{\tilde p}\right)^\frac{1}{\tilde p}=\left(\int_{\Sigma^n}d\mu_{g_t} |\phi|^{2\tilde p}\right)^\frac{2}{2\tilde p}=||\phi||^2_{2\tilde p}
  \end{equation*}
  so the choice $\tilde p = \frac p2 = \frac{n}{n-2}$ yields
    \begin{equation*}
\int_{\Sigma^n}d\mu_{g_t}\phi L_t \phi\geq\left(\frac{a_n}{K}-||R_t||_q\right)||\phi||^2_p \  
\end{equation*}
with $q = \frac{n}{2}$. Hence, the operator $L_t$ is positive if $(\frac{a_n} { K}-  ||R_t ||_q )>0$. Observe that for large enough $t$,
the metric $g_{ab}$ in the deformation region $2t<r<3t$ is asymptotically flat. Thus $g_{ab}(t) = \delta_{ab} + (1-\alpha_t(r)) h_{ab}$
where $h_{ab} = O(\frac{1}{r^{n-2}})$ in this region. Consequently, $|R_t| \leq \frac {\bar A}{r^n}$ in $2t<r<3t$ for some constant $\bar A$ independent of $t$. Thus 
\begin{equation}\label{curvenorm}
||R_t ||_q =\Bigl( \int _{{\Sigma }^n}d{\mu }_{ g_t} |R_t |^q  \Bigr) ^{\frac 1{q}} \leq 
\Bigl( \int _{2t}^{3t}d{\mu }_{ g_t} |R_t |^q  \Bigr) ^{\frac 1{q}} \leq
{\bar A} \Bigl( \int _{2t}^{3t}|{ \frac 1{r^n}}|^q r^{n-1}dr \Bigr) ^{\frac 1{q}}\leq 
\frac{B}{t^{\frac{n(q-1)} {q}}}
 \end{equation}
where the constants ${\bar A}$  and $B$ are independent of $t$. Therefore, there is some $t_0$ such
that for $t>t_0$, $L_t$ is a positive operator on $\phi\in C^\infty_0(\Sigma^n)$.  Hence Lemma  \ref{CLsolution} implies there exists a smooth positive solution
$G_t$ to 
\begin{equation}\label{cilt}
L_t G_t=-a_n{D_{t}}^2G_t+ {R_t}G_t=0
\end{equation}
for any $t>t_0$. From this point on,  $t$ will be taken  to have one fixed value above its lower bound. 

Note that, as  $g_{ab}(t)$ is flat for $r>3t$, the solution  $G_t$ of (\ref{cilt}) asymptotically has the standard expansion for a harmonic function
on flat space:
\begin{equation*}
{ G}_t=C_n\biggl(1+{\frac{A_0} {r^{n-2}}}+ \sum^\infty_{l=1}\sum_{ \kappa(l) }  \frac{A_{l\kappa }Y_{l\kappa}(\Omega )} {r^{l +n-2}}\biggr)
\end{equation*}
where $C_n$ is a dimension dependent constant determined by the normalization of the $n$-dimensional delta function and  $\kappa(l) $ is a set of integers that index the complete set of n-spherical harmonics $Y_{l\kappa}(\Omega)$ with principal Casimir indexed by  $l$.

Next, the manifold ${\Sigma }^n$ is compactified both topologically and geometrically. This compactification is motivated by the observation that stereographic projection of the round $n$-sphere results in ${\mathbb R}^n$ with flat metric and that $g_{ab}(t)$ is the flat metric for $r>3t$.  Inverting this procedure in the asymptotic region results in the closed manifold ${\tilde {\Sigma }^n}$ with metric $\tilde g_{ab}(t)$. 

Precisely, topologically compactify ${\Sigma }^n$ by attaching an n-ball to the asymptotic region using the trivial attaching map to form the closed manifold ${\tilde {\Sigma }^n}$. Define a smooth metric on 
${\tilde \Sigma ^n}$ by  ${\tilde g}_{ab}(t)={\phi }^{\frac{4}{n-2}}g_{ab}(t)$ where the conformal factor 
\begin{equation*}
\phi = \gamma +\frac{(1-\gamma)} {r^{n-2}}
\end{equation*}
 and
$\gamma (r)$ is a smooth bump function\footnote{For example, 
\begin{equation}\label{cfactor}
\gamma(r) =1- \frac {\int_{-\infty}^r dx' f(x'-4t)f(5t- x')}{\int_{-\infty}^\infty dx' f(x'-4t)f(5t - x')}
\end{equation}
 where $f$ is the smooth function
\begin{equation*}
f(x)=
\begin{cases}
0&x\leq 0\\
\exp(-\frac 1{x^2})& x>0\ .\\
\end{cases}
\end{equation*}
is such a bump function.
}
 which is $0$ for $r>5t$ and $1$ for $r< 4t$. Note that the form of the conformal factor yields ${\tilde g}_{ab}(t) = \frac{1} {r^{4}}\delta_{ab}$ for $r>5t$. The coordinate transformation $\bar r = \frac 1r$ in this region yields the flat metric $ds^2 = d{\bar r}^2 +{ \bar r}^2d\Omega^2_{n-1}$; thus ${\tilde g}_{ab}(t)$ is smooth everywhere in the neighborhood of $\bar r=0$. Denote this point $i$.

As ${\tilde g}_{ab}(t)$ is conformal to $g_{ab}(t)$, it follows from (\ref{cilt}) that  $\widetilde G_t = \phi^{-1}G_t$ with $\phi$ given by (\ref{cfactor}) is a positive solution of 
\begin{equation*}
-a_n{\tilde D}^2{\widetilde G}+ {\tilde R}{\widetilde G}=0
\end{equation*}
 on ${\tilde \Sigma} ^n-i$. Note that $\tilde D_a$ and $\tilde R$ are with respect to ${\tilde g}_{ab}(t)$. Furthermore, the expansion of $\widetilde G$ around  $i$ in the coordinate  ${\bar r}=\frac 1r$ is given by 
\begin{equation*}
{\widetilde G}=C_n\biggl( \frac 1 {{\bar r}^{n-2}} +{A_0} + \sum^\infty_{l=1}\sum_{ \kappa(l) }  A_{l\kappa }Y_{l\kappa}(\Omega ) {\bar r}^{l}\biggr)\ .
\end{equation*}
 It is manifestly apparent   that ${\widetilde G}$ is the Green's function
for the conformally invariant laplacian with respect to $\tilde g_{ab}(t)$, $\tilde L$, on ${\tilde {\Sigma ^n}}$, that is $\tilde L  {\widetilde G} =\delta_i$.

The final step is to show that there is a metric conformally related to $\tilde g_{ab}(t)$ on $\tilde \Sigma^n$ with everywhere positive scalar curvature. Let $\psi_0>0$ be the smooth positive solution (i.e. the ground state) to 
\begin{equation}\label{finalcil}
-a_n{\tilde D}^2{\psi _0}+ {\tilde R}{\psi _0}={\lambda }_0{\psi _0}
\end{equation}
on  $\tilde {\Sigma} ^n$.
Such a solution exists for  smooth  ${\tilde L}$ on a closed manifold \cite{Witt:2009za}.
Next, using $\psi_0$ as the conformal factor, construct the metric ${\hat g}_{ab}= {\psi _0}^{\frac 4 {n-2}}{\tilde g}_{ab}(t)$. 
The scalar curvature of ${\hat g}_{ab}$ is
\begin{equation}\label{finalcurv}
{\hat R}={\psi _0}^{-\bigl(\frac{n+2}{n-2}\bigr)}\bigl(-a_n{\tilde D}^2\psi _0 
+ {\tilde R}\psi _0\bigr)=\lambda _0\psi _0^{\bigl(1-\frac{n+2}{n-2}\bigr)} \ .
\end{equation} 
Integration of the left hand side of (\ref{finalcil}) against $\widetilde G$ yields
\begin{equation*}
\int_{\tilde \Sigma^n}d\mu_{\tilde g}\widetilde G \tilde L \psi_0 =\langle {\widetilde G}, 
\tilde L\psi _0\rangle= \lambda _0\langle {\widetilde G}, \psi _0 \rangle\ .\end{equation*}
However, 
\begin{equation*} \langle {\widetilde G}, 
\tilde L\psi _0\rangle=\langle\tilde L{\widetilde G},\psi _0\rangle= \langle \delta _{i},\psi _0
\rangle >0\ .
\end{equation*}
Thus $\lambda _0\langle {\widetilde G}, \psi _0 \rangle >0$ and, as $\langle {\widetilde G}, \psi _0\rangle$ is positive, it follows that ${\lambda }_0>0$.
Hence, by  (\ref{finalcurv}), ${\hat R}>0$ everywhere on $\tilde \Sigma^n$. Therefore, the Yamabe invariant is positive for $\tilde \Sigma^n$. By definition \ref{asymyambconst}, the asymptotically flat Yamabe invariant for $\Sigma^n$ is positive for the trivial compactification. Hence, the maximal Yamabe invariant for $\Sigma^n$ is also positive, ${\hat {\sigma }}(\Sigma ^n)>0$.

\end{proof}

Theorem \ref{compactify} readily generalizes to asymptotically flat manifolds with multiple asymptotic regions:

\begin{thm}\label{compactify2} Given an asymptotically flat manifold $\Sigma ^n$ with one or more asymptotic regions and asymptotically flat metric  $g_{ab}$ with
$R\geq 0$, then ${\hat {\sigma }}(\Sigma ^n)>0$.
\end{thm}
The proof of this theorem directly parallels that of theorem \ref{compactify} so will not be repeated here. The key difference is that quantities associated with the asymptotic regions will now be indexed and manipulations involving them typically involve sums. In particular, the parameter $t$ used in the deformation of the metric to one of zero scalar curvature becomes an indexed set of parameters $t_k$, one for each asymptotic region.  The bound on the norm of the deformed curvature (\ref{curvenorm}) then becomes a sum over the contributions from all asymptotic regions. Again, there will be a choice of $t_0$ such that if all $t_k>t_0$, then the conformal laplacian will be a positive operator.
Secondly, when $\Sigma^n$ is compactified, the added n-balls, behavior of the conformal factor on each asymptotic region and the poles of the Green's function will also be indexed by $k$. In particular, $\widetilde G$ will now be the Green's function for poles at $k$ points, $\tilde L\widetilde G = \sum_k \delta_{i_k}$, rather than one.  This replacement does not change either the result or conclusions. 

An immediate corollary of Theorem \ref{compactify2} is that the maximal Yamabe invariant characterizes an obstruction  to asymptotically flat initial data with a maximal slice on an asymptotically flat $n$-manifold:

\begin{cor}\label{maximalpos} Let $\Sigma ^n$ with metric $g_{ab}$ and extrinsic curvature $p_{ab}$ be  an  asymptotically flat initial data set with sources $\mu$ and $J^a$
that obey the dominant energy condition. If there is a maximal slice,  then $\hat \sigma(\Sigma ^n)>0$.
\end{cor}

\begin{proof} 
As $p=0$, the Hamiltonian constraint (\ref{hconstraint}) implies
\begin{equation*}
R=2\mu  +p_{ab}p^{ab}\geq 0
\end{equation*}
as $\mu \geq 0$ by the dominant energy condition. Theorem \ref{compactify2} now directly implies ${\hat {\sigma }}(\Sigma ^n)>0$.

\end{proof}
\noindent Hence, an asymptotically flat manifold $\Sigma^n$ with  $\hat \sigma(\Sigma ^n)\leq 0$ does not admit  asymptotically flat initial data with a maximal slice.

\section{Obstructions to global solutions of the Jang equation and a new singularity theorem for $(n+1)$-dimensional spacetimes} \label{sect5-jang}

We now prove the main results of this paper. First we prove that if  an asymptotically flat initial data set satisfying the dominant energy condition has a global solution to the Jang equation, then the Cauchy surface admits an asymptotically flat metric with zero scalar curvature. We then use this result to prove a new singularity theorem, that topological structures with nonpositive maximal Yamabe invariant collapse to form singularities.

Let ${\Sigma }^n$ with metric $g_{ab}$ and extrinsic curvature $p_{ab}$ be an asymptotically flat initial data set and let $\mu$ and $J^a$ be the corresponding sources. Form the new manifold $\Sigma^n\times{\mathbb R}$ with Riemannian metric given by line element
$ds^2 = g_{ab}dx^adx^b + d\tau^2$ where the coordinate $\tau$ is along ${\mathbb R}$.  The tensors $p_{ab}$, $\mu$ and $J^a$ are trivially extended so they are independent along parallel lines $\tau$. 
Let ${\cal G}^n_f\subset {\Sigma }^n \times \mathbb{R}$ 
be the graph of a function $f : {\Sigma }^n \to \mathbb{R}$ where ${\cal G}^n_f=\{(x,f(x))|x\in {\Sigma }^n \}$. 
The induced metric on ${\cal G}^n_f$ and its inverse are
 \begin{align*}{\bar g}_{ab}&=g_{ab}+D_{a}f D_{b}f\\
 {\bar g}^{ab}&=g^{ab} - \frac{D^{a}f D^{b}f}{1 + |Df|^2}\ .
 \end{align*}
  By construction, the mean curvature of the graph ${\cal G}^n_f$ is
$
H(f) = {\bar g}^{ab}\frac{D_{a} D_{b}f} {\sqrt{1 + |Df|^2}} 
% =\left(g^{ab} - \frac{D^{a}f D^{b}f}{1 + |Df|^2}\right)\left(\frac{D_{a} D_{b}f} {\sqrt{1 + |Df|^2}}  \right) 
$
 and the trace of $p_{ab}$ restricted to the graph is
$
P(f) = {\bar g}^{ab}p_{ab}%=\left(g^{ab} - \frac{D^{a}f D^{b}f}{1 + |Df|^2}\right) p_{ab}
$
The Jang equation is $H(f) = P(f)$; it can alternately be written as
\begin{equation*} \label{Jang's equation}
\left(g^{ab} - \frac{D^{a}f D^{b}f}{1 + |Df|^2}\right) \left(\frac{D_{a} D_{b}f} {\sqrt{1 + |Df|^2}} - p_{ab} \right) = 0.
\end{equation*}
By definition, when $f : {\Sigma }^n \to \mathbb{R}$ is a solution to the Jang equation, the graph ${\cal G}^n_f$
has mean curvature as prescribed by the trace of $p_{ab}$.

Schoen and Yau  proved the existence of solutions to the Jang equation   \cite{Schon:1981vd}. There are obstructions to finding global solutions, that is solutions with everywhere bounded $f$. If an obstruction occurs, then the initial data set contains apparent horizons, namely, closed manifolds ${\cal T}^{n-1} \subset {\Sigma }^n$ 
with $H_{{\cal T}^{n-1}} -P_{{\cal T}^{n-1}}=0$. In other words, the obstructions
are closed marginally outer trapped surfaces ${\cal T}^{n-1}$. Although \cite{Schon:1981vd} explicitly treats only the $n=3$ case, these results can be extended to dimensions $4$ and $5$ using the techniques found in \cite{schoen1975}. The results of  Eichmair allow the further extension of these results through dimension $7$ and imply a distributional solution in dimensions higher than $7$  \cite{Eichmair:2007}. As we assume the existence of  a regular global solution to the Jang equation in the our proof of Theorem \ref{jangpos}, it applies to Cauchy surfaces of dimension $n\leq 7$ though extension to higher dimensions may be possible. 

We now prove that  asymptotically flat initial data sets that have a global solution to the Jang equation must admit a metric of zero scalar curvature. 
\begin{thm}\label{jangpos} Let  $\Sigma ^n$ with metric $g_{ab}$ and extrinsic curvature $p_{ab}$ be an asymptotically flat initial data set with sources $\mu$ and $J^a$
that obey the dominant energy condition. If there is a global solution to the Jang equation, then there exists an 
asymptotically flat metric on $\Sigma ^n$ with $R=0$.
\end{thm}

\begin{proof}

Let $p_{ab}$ and $g_{ab}$ satisfy the constraint equations (\ref{hconstraint}) and (\ref{mconstraint}) and let $f$ be a function $f:\Sigma^n\to {\mathbb R}$. Define the new quantity $K = 1 + D_a f D^a f$; let
\begin{align*}
 \bar p_{ab} &= p_{ab} - \frac {D_aD_bf}{K^{\frac 12}} \\
 \bar g^{ab} &= g^{ab} - \frac {D^af D^bf}{K}\ .
 \end{align*}
In these variables, the Jang equation takes the form
\begin{equation}\label{jang}
\bar p_{ab} \bar g^{ab} =
%\left(p_{ab} - \frac {D_aD_bf}{K^{\frac 12}}\right)\left(g^{ab} - \frac {D^af D^bf}{K}\right) =  
0
\end{equation}
If $f$  is a global solution to the Jang equation, then the initial data $g_{ab}$,  $p_{ab}$ can be deformed into new  data $\bar g_{ab}$, $\bar p_{ab}$ on $\Sigma^n$  with zero prescribed mean curvature. The new data  does not necessarily satisfy the constraints, but 
does satisfy the  related equation (\ref{Rbarsat}), first derived in \cite{Schon:1981vd}; an alternate derivation follows below.

First note that the covariant derivative with respect to ${\bar g}_{ab}$ is related to that of ${g}_{ab}$ by
\begin{equation*}
(\bar D_b -D_b )v^c= C^a_{bc}v^c
\end{equation*}
for an any vector $v^c$ in $\Sigma^n$ 
where 
\begin{equation*}
C^a_{bc}= \frac 1K D^afD_bD_cf 
\end{equation*}
The Ricci curvature of ${\bar g}_{ab}$ is related to that of ${g}_{ab}$ by
\begin{align*}
\bar R_{ab} &= R_{ab} + Q_{ab}\nonumber\\
\bar R &= R + \bar g^{ab}Q_{ab} - \frac {D^af D^bf}{K} R_{ab}\\
Q_{ab}& = D_c C^c_{ab} - D_a C^c_{bc} + C^d_{ab}C^c_{dc} - C^d_{ac}C^c_{bd} \ .\nonumber
\end{align*}
Next, observe that
\begin{align*} 
p^2 &= 
%\left( \bar p_{ab} \bar g^{ab} + \frac {D^a f D^b f \bar p_{ab}}{K} + \frac {D^2 f}{K^{\frac 12}}\right)^2\nonumber\\&=  
\bar p_{cd} \bar g^{cd}  \left( \bar p_{ab} \bar g^{ab} + 2\frac {D^a f D^b f \bar p_{ab}}{K}\right)+
\left(\frac {D^a f D^b f \bar p_{ab}}{K}\right)^2\nonumber\\ 
&+ 2\frac {D^2 f \bar p_{cd} g^{cd}}{K^{\frac 12}}+ \frac {(D^2 f)^2}{K} \\
p_{ab}p_{cd}g^{ab}g^{cd} &= \bar p_{ab}\bar p_{cd}\bar g^{ac}\bar g^{bd} +2\frac{\bar g^{ab}D^c f  \bar p_{ca}D^d f  \bar p_{db}}{K} + 
\left(\frac {D^afD^bf \bar p_{ab}}{K}\right)^2\nonumber \\
& +2 \frac {D^a  D^b f \bar p_{ab}}{K^{\frac 12}} + \frac {D_a D_b f D^a D^bf }{K}
\end{align*}
Substituting these relations into the hamiltonian constraint yields
\begin{align}\label{Rbarpre}
\bar R -&\bar p_{ab}\bar p_{cd}\bar g^{ac}\bar g^{bd} -2\frac{\bar g^{ab}D^c f  \bar p_{ca}D^d f  \bar p_{db}}{K}
+  \bar p_{cd} \bar g^{cd}  \left( \bar p_{ab} \bar g^{ab} + 2\frac {D^a f D^b f \bar p_{ab}}{K}\right)\nonumber\\ 
&+ 2\frac {D^2 f \bar p_{cd} g^{cd}}{K^{\frac 12}}-2 \frac {D^a D^b f \bar p_{ab}}{K^{\frac 12}} \nonumber\\ 
&- \bar g^{ab}Q_{ab} + \frac {D^af D^bf}{K} R_{ab}  - \frac {D_a D_b f D^a D^bf }{K}+ \frac {(D^2 f)^2}{K}  = 2\mu
\end{align}
Next note that
\begin{align*}
\frac {D^a D^b f \bar p_{ab}}{K^{\frac 12}} &= D_a \left(\frac{ g^{ac}D^b f \bar p_{bc}}{K^{\frac 12}}\right) +
\frac{ D^c fD_aD_cf D^bf g^{ac}\bar p_{bc}}{K^{\frac 32}} - \frac {D^bf}{K^{\frac 12}}D_a (g^{ac}
 \bar p_{bc})\nonumber \\
&= \bar D_a\left (\frac {D^bf }{K^{\frac 12}}g^{ac} \bar p_{bc}\right) - \frac {D^bf}{K^{\frac 12}}D_a (g^{ac} \bar p_{bc})
\end{align*}
and
\begin{align*}
\frac {D^2  f \bar p_{cd}g^{cd}}{K^{\frac 12}} &= D_a \left(\frac{ D^a f \bar p_{cd}g^{cd}}{K^{\frac 12}}\right) +
\frac{ D^b fD_aD_b f D^a f \bar p_{cd}g^{cd}}{K^{\frac 32}} - \frac {D^af}{K^{\frac 12}}D_a (\bar p_{cd}g^{cd})\nonumber \\
&= \bar D_a\left (\frac {D^af }{K^{\frac 12}} \bar p_{cd}g^{cd}\right) - \frac {D^af}{K^{\frac 12}}D_a (\bar p_{cd}g^{cd} )
\end{align*}
and 
\begin{align*}
\frac {D^af}{K^{\frac 12}}D_a (\bar p_{cd}g^{cd} ) -\frac {D^bf}{K^{\frac 12}}D_a (g^{ac} \bar p_{bc})
 &= -\frac {D_af}{K^{\frac 12}}D_b (p^{ab}-pg^{ab}) 
 +\frac {D_a f}{K^{\frac 12}}D_b \left(\frac {D^aD^b f-D^2f g^{ab} }{K^{\frac 12}}\right)
 \nonumber\\
  &= -  \frac {D_af}{K^{\frac 12}}J^a  +\frac {D_a f}{K^{\frac 12}}D_b \left(\frac {D^aD^b f-D^2f g^{ab} }{K^{\frac 12}}\right)
\end{align*}
which allows the terms linear in ${\bar p}_{ab}$ to be replaced in (\ref{Rbarpre}) by a total divergence, terms involving the momentum 
$J^a$ and derivatives of $f$:
\begin{align}
\bar R - &\bar p_{ab}\bar p_{cd} \bar g^{ac}\bar g^{bd} - \frac {2 }{K}\bar g^{bd} D^a f \bar p_{ab} D^c f \bar p_{cd}+ \bar p_{ab}\bar g^{ab} \bar p_{cd}\left(\bar g^{cd}  + 2\frac{ D^cf D^df}{K}\right)\nonumber  \\
&- 2 \bar D_a \left( \frac {D^bf}{K^{\frac 12}} g^{ac}\bar p_{bc}- \frac {D^af}{K^{\frac 12}} g^{bc} \bar p_{bc}\right) +F= 2(\mu - \frac {D_a f J^a}{K^{\frac 12}})\label{partialhamiltonianc}\\
F&= - \bar g^{ab}Q_{ab} + \frac {D^af D^bf}{K} R_{ab}  - \frac {D_a D_b f D^a D^bf }{K}+ \frac {(D^2 f)^2}{K}  \nonumber\\
& -2\frac {D_a f}{K^{\frac 12}}D_b \left(\frac {D^aD^b f-D^2f g^{ab} }{K^{\frac 12}}\right)\label{Fterm}
\end{align}
Next, the divergence term can be rewritten in terms of ${\bar g}_{ab}$ instead of ${g}_{ab}$ using
\begin{align}\label{divchange}
\frac {D^bf}{K^{\frac 12}} g^{ac}\bar p_{bc}- \frac {D^af}{K^{\frac 12}} g^{bc} \bar p_{bc}
%&=
%\frac {D^bf}{K^{\frac 12}} g^{ac}\bar p_{bc}-\frac {D^af}{K^{\frac 12}}\left( \bar g^{bc} +\frac {D^bfD^cf}{K}\right) \bar p_{bc}\nonumber\\
&=\frac {D^bf}{K^{\frac 12}}\bar g^{ac}\bar p_{bc}-\frac {D^af}{K^{\frac 12}} \bar g^{bc} \bar p_{bc}
\end{align}
Finally, one  can show that $F$ vanishes; first  note that
\begin{align*} g^{ab}Q_{ab}&=  \frac{(D^2 f)^2}{K} +  \frac{D^a f (D_aD^2 f-D^2 D_a f)}{K} -  \frac{D_a D_b f D^a D^b f}{K}\\&- \frac {D^afD^bf D_aD_b f D^2f}{K^2} + \frac {D^af D^bf D_a D_cf D_b D^cf}{K^2}\\
-\frac {D^a f D^bf} {K}Q_{ab}&=\frac {D^af D^bf D_a D_cf D_b D^cf}{K^2}- \frac {D^afD^bf D_aD_b f D^2f}{K^2} 
\end{align*}
so that
\begin{align}\label{Qterm} - \bar g^{ab}Q_{ab}&= - \frac{(D^2 f)^2}{K} - \frac{D^a f (D_aD^2 f-D^2 D_a f)}{K} +\frac{D_a D_b f D^a D^b f}{K}\nonumber\\&+2\frac {D^afD^bf D_aD_b f D^2f}{K^2} -2\frac {D^af D^bf D_a D_cf D_b D^cf}{K^2}
\end{align}
Next, one finds that
\begin{align} \label{derof}
 -2\frac {D_a f}{K^{\frac 12}}D_b \left(\frac {D^aD^b f-D^2f g^{ab} }{K^{\frac 12}}\right)&= 
2\frac {D^a f\left( { D_a D^2 f-D^2 D_a f }\right) }{K}\nonumber\\
&-2 \frac {D^afD^bf D_aD_b f D^2f}{K^2} +2 
\frac {D^af D^bf D_a D_cf D_b D^cf}{K^2}\end{align}
Hence, substituting (\ref{Qterm}) and (\ref{derof}) into  (\ref{Fterm}) yields
\begin{align*}
F&= \frac{D^a f (D_aD^2 f-D^2 D_a f)}{K} + \frac {D^af D^bf}{K} R_{ab}\\&= 
- \frac {D^af D^bf}{K} R_{ab}+ \frac {D^af D^bf}{K} R_{ab}=0
\end{align*}
Using this and (\ref{divchange}), the transformed Hamiltonian constraint (\ref{partialhamiltonianc}) can be written
\begin{align*}%\label{Rbarnot}
\bar R - &\bar p_{ab}\bar p_{cd} \bar g^{ac}\bar g^{bd} - \frac {2 }{K}\bar g^{bd} D^a f \bar p_{ab} D^c f \bar p_{cd}+ \bar p_{ab}\bar g^{ab} \bar p_{cd}\left(\bar g^{cd}  + 2\frac{ D^cf D^df}{K}\right) \nonumber\\
&- 2 \bar D_a \left( \frac {D^bf}{K^{\frac 12}} \bar g^{ac}\bar p_{bc} -\frac {D^af}{K^{\frac 12}}\bar g^{cd}\bar p_{cd}\right) = 2(\mu - \frac {D_a f }{K^{\frac 12}}J^a)\end{align*}
where $\bar R$ is the scalar curvature of ${\bar g}_{ab}$ $\bar D_a$ is the covariant derivative with respect to ${\bar g}_{ab}$.
As $f$ is assumed to satisfy the Jang equation, (\ref{jang}), this simplifies to
\begin{equation}\label{Rbarsat}
\bar R   - \frac {2 }{K}\bar g^{bd} D^a f \bar p_{ab} D^c f \bar p_{cd}
- 2 \bar D_a \left( \frac {D^bf}{K^{\frac 12}} \bar g^{ac}\bar p_{bc} -\frac {D^af}{K^{\frac 12}}\bar g^{cd}\bar p_{cd}\right) = 2(\mu - \frac {D_a f }{K^{\frac 12}}J^a)+\bar p_{ab}\bar p_{cd} \bar g^{ac}\bar g^{bd} \end{equation}
where $\bar R$ is the scalar curvature and $\bar D_a$ is the covariant derivative with respect to ${\bar g}_{ab}$.  The right hand side is nonnegative if  the matter source satisfies the dominant energy condition
as $ | \frac {D_a f }{K^{\frac 12}}|\leq 1$. Hence $\bar R$ satisfies the inequality
\begin{equation*}%\label{Rbarsat2}
\bar R - \frac {2 }{K}\bar g^{bd} D^a f \bar p_{ab} D^c f \bar p_{cd}
- 2 \bar D_a \left( \frac {D^bf}{K^{\frac 12}} \bar g^{ac}\bar p_{bc} -\frac {D^af}{K^{\frac 12}}\bar g^{cd}\bar p_{cd}\right) \geq 0\end{equation*}
Multiplication by a function $\phi^2$  and rearrangement yields
\begin{align*}
\phi^2 \bar R &+2 \bar g^{bd}D_b\phi D_d \phi - 4\phi D_a \phi \frac {D^af}{K^{\frac 12}}\bar g^{cd}\bar p_{cd}
\nonumber\\
&- 2 \bar D_a \left( \phi^2 \frac {D^bf}{K^{\frac 12}} \bar g^{ac}\bar p_{bc} -\phi^2 \frac {D^af}{K^{\frac 12}}\bar g^{cd}\bar p_{cd}\right) \geq 2 \bar g^{bd}\left( \phi \frac { D^a f  }{K^{\frac 12}}\bar p_{ab} -D_b \phi\right) \left( \phi \frac { D^c f  }{K^{\frac 12}}\bar p_{cd} -D_d \phi\right)   \end{align*}
which, again assuming the the Jang equation is satisfied, simplifies  to
\begin{align} \label{phijang}
\phi^2 \bar R +2(\bar D \phi)^2 - 2 \bar D_a \left( \phi^2 \frac {D^bf}{K^{\frac 12}} \bar g^{ac}\bar p_{bc} \right) \geq 0 \end{align}
where $(\bar D \phi)^2 =\bar g^{bd}\bar D_b\phi \bar D_d \phi=\bar g^{bd}D_b\phi D_d \phi $.
Since the solution is global, the graph ${\cal G}^n_f={\Sigma }^n$.  Integration  of (\ref{phijang}) over the graph therefore yields
\begin{align*}
\int _{{\Sigma }^n}d{\mu }_{\bar g}(\phi^2 \bar R +2(\bar D \phi)^2 )
- 2 \int _{{\Sigma }^n}d{\mu }_{\bar g} \bar D_a \left( \phi^2 \frac {D^bf}{K^{\frac 12}} \bar g^{ac}\bar p_{bc} \right)\geq 0 
\end{align*}
If $\phi \in C^{\infty }_0({\Sigma }^n)$, then the integral over the divergence vanishes as there are no boundaries interior to the asymptotic regions. Therefore,
\begin{align*}
\int _{{\Sigma }^n}d{\mu }_{\bar g}(\phi^2 \bar R +2(\bar D \phi)^2)  \geq 0 \ .\end{align*}
This implies,
as  $a_n={\frac{4(n-1)} {n-2}}> 2$ for $n\geq 3$,
\begin{align*}%\label{positiveop}
\int _{{\Sigma }^n}d{\mu }_{\bar g} \phi\bar L\phi=
\int _{{\Sigma }^n}d{\mu }_{\bar g} (a_n(\bar D \phi)^2 +  \phi^2 \bar R )
 \geq \int _{{\Sigma }^n}d{\mu }_{\bar g} (2(\bar D \phi)^2 +  \phi^2 \bar R  ) \geq 0 \ .\end{align*}
Thus the conformal laplacian operator (\ref{conflaplacian}) for metric $\bar g_{ab}$ is positive on $\phi \in C^{\infty }_0({\Sigma }^n)$.
Therefore, by Theorem \ref{conftoR=0} there is a smooth, everywhere positive solution $\phi$ on $\Sigma^n$ where
$\phi\to 1$ with asymptotically flat fall-off as $r\to \infty$ in each asymptotic region.
Define the metric ${\tilde g}_{ab}={\phi }^{\frac{4} {n-2}}{\bar g}_{ab}$. This metric is a complete, asymptotically flat metric
with zero scalar curvature everywhere.

\end{proof}

This result directly implies a new singularity theorem for manifolds with nonpositive maximal Yamabe invariant. A spacetime satisfies the {\it null convergence condition}, also known as the {\it null energy condition}  if $R_{ab} W^a W^b \geq 0$ 
for all null $W^a$, the {\it weak energy condition} if $T_{ab} W^a W^b \geq 0$ 
for all timelike $W^a$  and the {\it dominant energy condition}  if the weak energy condition holds and
$T_{ab} W^b T^a_{c} W^c \leq 0 $ \cite{HawkingEllis}.  Notice the least restrictive of these energy conditions 
is the null convergence  condition. If the dominant energy condition is satisfied,  then so is the null convergence condition by a continuity argument. The null convergence condition is  the energy condition required in the singularity theorem of Penrose  \cite{Penrose:1964wq,HawkingEllis3}. If a spacetime is the maximal evolution of an initial data set satisfying the dominant energy condition, it automatically satisfies  the null convergence condition. However, for generality, the theorem below assumes the null convergence condition in the spacetime separately from the assumption of the dominant energy condition on the Cauchy surface. 

\begin{thm}\label{jangsing} If  a spacetime $M^{n+1}$ has a Cauchy surface $\Sigma^n$   with  ${\hat {\sigma }}(\Sigma ^n)\leq 0$, asymptotically flat initial data  and
sources 
that obey the dominant energy condition, then $M^{n+1}$ contains one or more apparent horizons. Hence if $M^{n+1}$ satisfies the null convergence condition, then it is null geodesically incomplete.
\end{thm}

\begin{proof}
Assume that a global solution of the Jang equation exists for the asymptotically flat initial data set on $\Sigma^n$. By Theorem \ref{jangpos},
it follows that $\Sigma^n$ admits an asymptotically flat metric with zero scalar curvature. Theorem \ref{compactify2} then implies that 
 ${\hat {\sigma }}(\Sigma ^n)>0$, in contradiction to the assumption that ${\hat {\sigma }}(\Sigma ^n)\leq 0$.
Consequently ${\Sigma ^n}$ must not admit a global solution to to the Jang equation. Hence, as obstructions to a global solution are closed submanifolds ${\cal T}^{n-1} \subset {\Sigma }^n$ with $H_{{\cal T}^{n-1}} -P_{{\cal T}^{n-1}}=0$,  the initial data set must contain apparent horizons. This immediately implies that $M^{n+1}$ is singular by the Penrose singularity theorem.\footnote{ Note that it suffices that the trapped surfaces be outer trapped surfaces in the proof of the singularity theorem. See, for example \cite{Friedman:1993ty}.} 
\end{proof}
\noindent The apparent horizons forming the obstruction to the Jang equation can be either future or past trapped (or both). Therefore, the singularities may be in either the past and/or future evolution.

 The key feature of Theorem \ref{jangsing} is the proof of the existence of one or more apparent horizons.  Consequently, other singularity theorems with conditions requiring an apparent horizon hold for spacetimes with Cauchy surface $\Sigma^n$   with  ${\hat {\sigma }}(\Sigma ^n)\leq 0$, asymptotically flat initial data  and
sources 
that obey the dominant energy condition. In particular, such spacetimes are singular if they also satisfy the strong energy condition and generic condition by the Hawking-Penrose theorem   \cite{Hawking:1969sw,HawkingEllis4}:
 \begin{thmH} Spacetime ${\cal M}$ with metric $\gamma_{ab}$ is not timelike and null geodesically complete if
 the chronology condition holds, the strong energy condition and the generic  condition are satisfied, and there exists one of the following:
1) a compact achronal set without edge,
2) a closed trapped surface,
3)  a point $p$ from which every past directed (or future directed) null geodesic has null expansion that becomes negative.
\end{thmH}
\noindent 
The strong energy condition is that $R_{ab} W^aW^b\geq 0$ for non-spacelike $W^a$. The generic condition is that every non-spacelike geodesic contains a point at which $K_{[a}R_{b]cd[e}K_{f]}K^cK^d\neq 0$ where $K^a$ is the tangent to the geodesic.
 The generic condition is satisfied for the case of vacuum spacetime containing gravitational radiation and in many other physical situations.  

\section{Asymptotically flat simply connected $4$-manifolds with nonpositive maximal Yamabe invariant }\label{sect6}

In $5$ or more spacetime dimensions, there are an infinite number of asymptotically flat spacetimes with topologically distinct, simply connected Cauchy surfaces. 
Consequently Theorem \ref{jangsing} implies collapse of a set of topological structures not addressed by prior singularity theorems. Given this, it is useful to discuss the construction and characterization of such spacetimes. We do so below, concentrating on the interesting case of $5$ dimensional asymptotically flat spacetimes with  Cauchy surfaces $\Sigma^4$.

Closed manifolds with obstructions to positive scalar curvature can be used to construct asymptotically flat manifolds with nonpositive maximal Yamabe invariant. Recall that smooth compactification of an asymptotically flat manifold $\Sigma^n$ results in a  smooth closed manifold $\widetilde \Sigma^n$. By Definition \ref{asymyamb}, if $\widetilde \Sigma^n$ has nonpositive Yamabe invariant, the corresponding asymptotically flat manifold has $\hat \sigma(\Sigma^n)\leq 0$. Consequently, by puncturing  closed manifolds $\widetilde \Sigma^n$ with obstructions to positive scalar curvature, one can construct asymptotically flat manifolds $\Sigma^n$ with  $\hat \sigma(\Sigma^n)\leq 0$. Furthermore, obstructions to positive scalar curvature are well known to be related to topological properties; therefore topological and smooth invariants of closed manifolds characterize classes of asymptotically flat manifolds with
$\hat \sigma(\Sigma^n)\leq 0$.

In $4$ dimensions,  topological obstructions to positive curvature are characterized by the $\widehat A$-genus. In addition, the differentiable structure of the manifold can also produce an obstruction to positive curvature. The Seiberg-Witten invariants characterize obstructions produced by both topology and differentiable structure.

\subsection{Nonpositive maximal Yamabe invariant from topological obstructions in $4$ dimensions}

 If $M^4$ is a smooth spin manifold, i.e. a spin manifold that admits a differentiable structure, the $\widehat A$-genus can be defined in terms of the index of the Dirac operator. As the dimension is even, a complex spin bundle $S$ over $M^4$ has a natural decomposition into the sum  $S_+\oplus S_-$, the eigenspaces of the complex volume element  $\omega_{\mathbb C}$ of the Clifford algebra.
Let $D^{\, +}$ be the restriction of the Dirac operator $D$ to $S_+$ and $D^{\, -}$ corresponding restriction  to $S_-$. 
Note $D^ {\, +}:S_+\to S_-$ and the adjoint of  $D ^{\, +}$ is $D^{\, -}$. 
The index of $D^{\, +}$,  $ind(D^{\, +})= dim(\,ker\, D^{\, +}) - dim(\, ker\, D^{\, -})$.  This invariant  
is the $\widehat A$-genus, $ind(D^{\, +})=\widehat A(M^4)$. 

The Weitzenb\"ock formula for the Dirac laplacian is 
\begin{equation*}
D^{\, 2}\psi =-\nabla ^2\psi +\frac{1}{4}R\psi. 
\end{equation*}
If $M^4$ admits a metric with $ R>0$, this implies, as the operator $-\nabla ^2 +\frac{1}{ 4}R$ is positive, that $ker\, D^{\, 2}=0$. Consequently, as $ker\, D^{\, +} + ker\, D^{\, -}= ker\, D=ker\, D^{\, 2}$, $ind(D^{\, +})=\widehat A(M^4)=0$; the $\widehat A$-genus vanishes. Therefore, if a closed  $4$-manifold has nonvanishing $\widehat A$-genus, it does not admit a metric of positive scalar curvature. This result is due to Lichnerowicz \cite{lichnerowicz}.

This result extends to smooth, asymptotically flat $4$-manifolds:
\begin{lem}\label{lichvanish} A closed smooth spin manifold $M^4$ with $\widehat A (M^4)\neq 0$ has nonpositive Yamabe invariant,  $\sigma (M^4)\leq 0$. Furthermore the maximal Yamabe invariant of the asymptotically flat manifold $M^4-S$, $S$  a finite set of points, is also nonpositive,  ${\hat \sigma }(M^4-S)\leq 0$.
\end{lem}
\begin{proof}
As
$\widehat A (M^4) \neq 0$, it follows that $R \leq 0 $ for all metrics on $M^4$.  Hence, $\sigma (M^4)\leq 0$.
Next, observe that by Theorem \ref{trivialunique}, the compactification of   $M^4-S$ 
to $M^4$ is unique. Consequently, by Definition \ref{asymyamb},
${\hat \sigma }(M^4-S)\leq 0$.
\end{proof}

In general, the $\widehat A$-genus for any closed $4$-manifold can be computed in terms of its signature.
 When $n=4k$, $k\geq 1$  the signature of a closed n-manifold is that of the quadratic form
\begin{equation*}%\label{quadform}
Q:H^{2k}(M^n;{\mathbb Z})\otimes H^{2k}(M^n;{\mathbb Z})\rightarrow {\mathbb Z}
\end{equation*}
where $Q(\alpha , \beta )= (\alpha \cup \beta )[M^n] $.\footnote{In terms of the de Rham cohomology over 
${\mathbb R}$, 
$Q(\alpha , \beta ) =\int _{M^n}{\alpha }\wedge {\beta }
$
for $\alpha , \beta \in H^{2k}( M^n; {\mathbb R})$. }
Then  $\widehat A(M^n)=-\frac 1 8\hbox{\rm sig}(M^n)$. 
If $M^4$ is a smooth closed $4$-manifold, 
$\widehat A[M^4]=-{\frac 1 8}\tau (M^4)$ 
where $\tau (M^4)$ is 
the Hirzebruch signature,
\begin{equation*}
\tau (M^4) = \frac 1{48 \pi^2}\int _{M^4}C_{abcd} ^* C^{abcd} d\mu _g\  
\end{equation*}
with $C_{abcd}$ being the Weyl curvature of any riemannian metric $g$ on $M^4$. Note that the Hirzebruch signature can be computed for any smooth $4$-manifold, with or without spin structure.

The intersection form of simply connected $4$-manifolds is particularly well understood. It is unimodular and its basic building blocks are $H=\left( \begin{smallmatrix} 0&1\\ 1&0 \end{smallmatrix} \right)$, $ \langle 1\rangle$, $ \langle -1\rangle$ and $E_8$ \cite{GompfStipsicz}.  Furthermore, it classifies simply connected closed $4$-manifolds in the following sense according to the results of Freedman \cite{Freedman}:
For every unimodular symmetric bilinear form $Q$ there exists a simply connected closed topological $4$-manifold $M$ such that
$Q_M\cong
 Q$. If $Q$ is even, this manifold is unique up to homeomorphism. If $Q$ is odd, there are exactly two different homeomorphism types of manifolds with the given intersection form. At most one of these carries a smooth structure. Consequently, simply connected smooth $4$-manifolds are determined up to homeomorphism by their intersection forms. 
 
Secondly, the quadratic form of the connected sum of two simply connected $4$-manifolds is the direct sum of their quadratic form, $Q_{M_1\#M_2} = Q_{M_1}\oplus Q_{M_2}$. Consequently, families of simply connected $4$-manifolds can be constructed by taking connected sums of a set of fundamental building blocks, $S^4$, $S^2\times S^2$, ${\mathbb C}P^2$, $\overline{{\mathbb CP}^2}$,$K3$ and $E8$ with quadratic forms
 $Q_{S^4}=0$, $Q_{S^2\times S^2}=H$,  $Q_{K3}=2(-E_8)\oplus 3H$, $Q_{{\mathbb C}P^2}= \langle 1\rangle$,  $Q_{\overline{{\mathbb C}P^2}}=  \langle -1\rangle$ and $Q_{E8}=E_8$. As the quadratic forms of $S^4$, $S^2\times S^2$, $K3$ and $E8$ are even, these manifolds have vanishing second Stiefel-Whitney class and admit a spin structure;  as the quadratic forms of ${\mathbb C}P^2$ and $\overline{{\mathbb C}P^2}$ are odd, they do not.  In addition, Rokhlin's theorem  \cite{Rokhlin} states that the signature of any closed  smooth spin $4$-manifold  is a multiple of
16. A computation of its signature demonstrates that $E8$ is a  spin $4$-manifold which 
admits no smooth structure.
 \footnote{Alternately, the Hirzebruch signatures are $\tau (S^4)=0$, 
$\tau ({\mathbb C}P^2) = 1$,  $\tau (S^2\times S^2)=0$ and $\tau (K3)= 16$. Consequently, Rokhlin's theorem \cite{Rokhlin} implies that $CP^2$ is a closed smooth $4$-manifold which does not admit a spin structure.}

We now apply these results to exhibit an infinite family of simply connected, asymptotically flat $4$-manifolds with nonpositive maximal Yamabe invariant. First, as the signature of $K3$ is $16$, its $\widehat A$-genus is nonzero.
It follows by Lemma \ref{lichvanish}, that the asymptotically flat manifold obtained by removing a point $p$, $K3 - p$, has  ${\hat \sigma }(K3-p)\leq 0$. 
Connected sums of $K3$ with itself and $S^2\times S^2$ produce more examples. For example, the connected sum $K3\# K3$  also has nonzero $\widehat A$-genus and consequently the asymptotically flat manifold $K3\# K3-p$ has nonpositive maximal Yamabe invariant. The manifold $K3\# (S^2\times S^2)-p$ similarly also has nonpositive maximal Yamabe invariant.  Furthermore,  the quadratic form of a smooth simply connected spin $4$-manifold $M^4$ is homeomorphic to $2kE8 \oplus nH$ $k,n$ integers so long as $b_2(M^4) \geq \frac {11}{8} | \tau(M^4)|$ where $b_2(M^4)$ is the second betti number of $M^4$. Therefore there is an infinite set of smooth simply connected spin $4$-manifolds with nontrivial signature and consequently a corresponding infinite set of asymptotically flat manifolds with nonpositive maximal Yamabe invariant.

\subsection{Nonpositive maximal Yamabe invariant and the Seiberg-Witten invariants}

In contrast to higher dimensional manifolds, closed $4$-manifolds can admit a countably infinite number of distinct differentiable structures. Furthermore, open $4$-manifolds can admit an uncountably many distinct differentiable structures as dramatically illustrated for ${\mathbb R}^4$. Simply connected $4$-manifolds also carry an countably infinite number of distinct differentiable structures; a theorem of Friedman and Morgan shows that simply connected manifolds corresponding to the intersection forms $2n(-E_8)\oplus (4n-1)H$, $n\geq1$, and
$(2k - 1)\langle 1 \rangle \oplus N\langle -1 \rangle$, $k\geq 2$, $N\geq 10k-1$, each carry infinitely many distinct differentiable structures \cite{FM, GompfStipsicz}. Those with intersections forms of the first type have nonzero $\widehat A$-genus and consequently collapse from their topology. However, those of the second type, realized by certain connected sums of ${\mathbb C}P^2$ and $\overline{{\mathbb C}P^2}$ do not. However, certain differentiable structures on such manifolds also produce an obstruction to positive curvature. These obstructions can be characterized by  the existence of solutions to the Seiberg-Witten equations and are  thus related to the Seiberg-Witten invariants.  They can be viewed as arising from the topological structure of an associated bundle, the complex spin bundle, over the $4$-manifold.  Consequently, as discussed below, the related set of asymptotically flat manifolds will have nonpositive maximal Yamabe invariant. 

A complex spin structure on a closed $4$-manifold $M^4$ is given by replacing 
the group $Spin(4)$ with the group $Spin_{\mathbb C}(4)$. Clearly, all 
$4$-manifolds which admit a spin structure automatically admit a complex spin structure. More generally,  all non-spin $4$-manifolds also admit a complex spin structure by a theorem of Wu.
Let $D_A$ be the dirac operator associated with the $U(1)$ connection $A$ on a given bundle of $Spin_{\mathbb C}(n)$ spinors $W$  on $M^4$. As in the spinor case, $W$ can be decomposed into the sum of eigenspaces
$W_+\oplus W_-$ in even dimensions with 
 $D_A^{\, +}$ the restriction of $D_A$ to $W_+$ and $D_A^{\, -}$ the corresponding restriction  to $W_-$. Note
$D_A^ {\, +}:W_+\to W_-$ and the adjoint of  $D_A ^{\, +}$ is $D_A^{\, -}$. 
The Weitzenb\"ock formula for the $Spin_{\mathbb C}(n)$ Dirac operator is given by
\begin{equation*}
D^{\, 2}_A\psi =-\nabla ^2\psi +\frac 14 R\psi +\frac 12 F_A\psi . 
\end{equation*}
This expression now contains both the scalar curvature $R$ of the riemannian metric and the curvature
$F_A$ of the connection $A$. 
 Consequently, the index of $D_A$  yields information regarding obstructions to
scalar curvature; however, the existence of an obstruction now also depends on the curvature of the connection $A$.  

In $4$ dimensions, the  Seiberg-Witten equations provide a particularly fruitful choice of connection and curvature. 
The generalized Seiberg-Witten equations  are
\begin{equation*}%\label{SeibWitt}
D_A\uppsi =0\ \ \ \ {\rm and}\ \ \ \  F^+ _A= q (\uppsi ) +i\omega
\end{equation*}
where $D_A$ is the Dirac associated to $A$, $F^+ _A$ is self-dual part of the curvature 2-form of $A$,
$q $ is the map from $W_+$ to imaginary self-dual 2-forms which takes the square of the spinor $\uppsi$, namely
$q (\uppsi )= \uppsi \otimes \uppsi^* -\frac{I|\uppsi |^2}{ 2 }$, and $\omega $ is a real self-dual 2-form, typically 
chosen to be zero or harmonic. The Seiberg-Witten equations
are the above equations with $\omega \equiv 0$. The case of $\omega \not\equiv 0$ corresponds to
a perturbation of the Seiberg-Witten equations needed to avoid singular points
in the moduli space.
Solutions to the Seiberg-Witten equations are called {\it monopoles} as these equations 
are the field equations of massless magnetic monopoles on the manifold $M^4$. 

Given a closed $4$-manifold $M^4$ with a $U(1)$ gauge field, the space
${\cal A }[M^4]$ is the space of pairs $(A,\uppsi )$ with $A$ a $U(1)$ connection on the complex line bundle $L$
and $\uppsi \in W_+$. Moreover, the gauge transformations are given by smooth maps of
$M^4$ into $U(1)$, ${\cal G}[M^4]=C^\infty (M^4,U(1))$.
The moduli space is ${ {\cal B}}[M^4]={\cal A }[M^4]/{\cal G}[M^4]$ and the irreducible moduli space is
${ {\cal B}^*}[M^4]={\cal A }^*[M^4]/{\cal G}[M^4]$ where ${\cal A }^*[M^4]\subset {\cal A }[M^4]$ is
the subspace of configurations with $\uppsi \not \equiv 0$. Instead of using the full group of gauge transformations,
one can fix a base point on $x_0\in M^4$ and consider the group of base point fixing gauge transformations 
${\cal G}_0[M^4]=\{ g\in {\cal G}[M^4]| g(x_0)=id \}$.
Note that ${\cal G}[M^4]/{\cal G}_0[M^4]\cong U(1)$. The advantage of this group is that it acts freely on gauge 
configurations. One can define the corresponding moduli spaces 
\begin{equation*}{{\tilde {\cal B}}}[M^4]={\cal A }[M^4]/{\cal G}_0[M^4]\  {\rm and}\  {{\tilde {\cal B}^*}}[M^4]={\cal A }^*[M^4]/{\cal G}_0[M^4]\ .\end{equation*}
There is a $U(1)$ bundle which relates the moduli spaces defined in terms of ${\cal G}_0[M^4]$ to those defined in terms of
${\cal G}[M^4]$, namely, ${{\tilde {\cal B}}}[M^4]={\cal A }[M^4]/{\cal G}_0[M^4]$ is the total space of a $U(1)$ bundle over
${{\tilde {\cal B}}}[M^4]={\cal A }[M^4]/{\cal G}[M^4]$. The same is true of ${{\tilde {\cal B}^*}}[M^4]$  and ${ {\cal B}^*}[M^4]$.

The monopole moduli space is
\begin{equation*}{\cal M}_{L }[M^4]=\{ (A,\uppsi )\in { {\cal B}}[M^4]|D^+_A\uppsi =0, \ \ F^+_A = \sigma (\uppsi ) \}\ .\end{equation*}
\noindent Observe that ${\cal M}_{L }[M^4]\subset { {\cal B}}[M^4]$ is a finite dimensional subspace. The moduli space depends 
on the choice of $L$. One 
problem is the moduli space ${\cal M}_{L }[M^4]$ can be singular. This may occur for several different reasons. First, the moduli space
can be singular due to fixed points from the action of gauge transformations. Moreover, reducible solutions 
(ones with $\uppsi \equiv 0$) also give points for which the gauge group does not act freely. Finally, the equations themselves
may not satisfy the conditions of the implicit function theorem which is needed to prove the desired properties of the
moduli spaces. Therefore it is useful to define the following moduli spaces.

The generalized or the perturbed monopole moduli space is
\begin{equation*}{\cal M}_{L, \omega }[M^4]=\{ (A,\uppsi )\in { {\cal B}}[M^4]|D^+_A\uppsi =0, \ \ F^+_A = \sigma (\uppsi ) +i\omega \}\end{equation*}
where the use of ${\cal B}$ means that all gauge transformations are removed. Similarly, removing only the base point fixing gauge transformations, one can define 
\begin{equation*}{\tilde {\cal M}}_{L, \omega }[M^4]=\{ (A,\uppsi )\in { {\tilde {\cal B}}}[M^4]|D^+_A\uppsi =0, \ \ F^+_A = \sigma (\uppsi ) +i\omega \}\ .\end{equation*}
The monopole moduli spaces ${\tilde {\cal M}}_{L, \omega }[M^4]$ and ${\cal M}_{L }[M^4]$ are also related via $U(1)$ bundles. 
 They are finite dimensional because they are solutions of 
nonlinear elliptic equations; however, these spaces can be shown to be compact. Moreover, they can shown to be smooth 
manifolds.

The Seiberg-Witten $SW(L)$ invariants are defined to be zero if $dim ({\cal M}_{L,\omega}[M^4])$ is odd; otherwise
\begin{equation*}SW(L)= \int_{{\cal M}_{L,\omega }[M^4] } c_1^d\end{equation*}
\noindent where $d= {\frac12}dim ({\cal M}_{L,\omega}[M^4])$ and $c_1$ is the first Chern class in $H^2({\cal B^*}[M^4];{\mathbb Z})$ 
corresponding to the $U(1)$ bundle of ${{\tilde {\cal B}^*}}$ over ${{ {\cal B}^*}}$. 
One can show
that if the Seiberg-Witten invariant is well defined, then it does not depend on $\omega $. Note that the Seiberg-Witten invariant depends both on the smooth structure of the closed $4$-manifold and on the choice 
of complex spin structure. In particular, one can show that $SW(L)$
is well defined when it satisfies the conditions of the next theorem. In addition, it is clear that
if $SW(L)\neq 0$, one has solutions to Seiberg-Witten equations.

As Seiberg-Witten invariants are diffeomorphism invariant, obstructions to positive curvature 
characterized by them can arise from the differentiable structure of the manifold.

\begin{thm}\label{diffcolapse} If a closed orientable smooth $4$-manifold $M^4$ with vanishing first betti number
admits a solution to the Seiberg-Witten equations and has either $[c_1(L)]^2[M^4]\geq 0$ or $b_2^+(M^4)>1$
then it has nonpositive Yamabe invariant, $\sigma(M^4)\leq 0$. Consequently, the maximal Yamabe invariant of  
the asymptotically flat manifold $M^4-S$, $S$ a finite set of points, is also nonpositive, ${\hat {\sigma }}(M^4-S )\leq 0$.

\end{thm}

\begin{proof}
Assume $M^4$  has a solution $(A,\uppsi )$ to the Seiberg-Witten equations. Application of the Weitzenb\"ock 
formula followed by use of the Seiberg-Witten equations implies
\begin{align*}
0=\int _{M^4}\uppsi ^* D^{\, 2}_A\uppsi \ d\mu& =\int _{M^4} (-\uppsi ^*\nabla ^2\uppsi +\frac 14 \uppsi ^*R \uppsi +
\frac 12 \uppsi ^*F_A\uppsi )\ d\mu \nonumber \\
&=\int _{M^4} (-\uppsi ^*\nabla ^2\uppsi +\frac 14 \uppsi ^*R \uppsi + \frac 12 \uppsi ^*F^+_A\uppsi )\ d\mu 
= \int _{M^4} (|\nabla\uppsi |^2+\frac 14 R |\uppsi  |^2 +\frac 14| \uppsi |^4)\ d\mu \ . 
\end{align*}
Hence,  if $R> 0$, then the only possible solutions to the Seiberg-Witten equations are $F^+ _A=0$ with $\uppsi \equiv 0$. 
Note that every connection has a split in terms of self-dual and anti self-dual connection; furthermore $F _A=F^+ _A+F^- _A$. Moreover, 
$F _A$ represents a cohomology class and may be chosen to be a harmonic real form. If $F^+ _A=0$,
then $F _A=F^- _A$. The riemannian geometry question reduces to a question of $U(1)$ instantons over  $M^4$; specifically
 if the manifold does not admit an anti-self dual $U(1)$ connection, then $M^4$ admits no metric with 
$R\geq 0$.

If $[c_1(L)]^2[M^4]\geq 0$, note that
every $U(1)$ bundle has unique bundle connection with harmonic curvature corresponding to the 
its first Chern class, $c_1(L)={\frac 1 {2\pi }}F _A$.
The first Chern class squared $[c_1(L)]^2[M^4]$ of the line bundle $L$ for the associated complex 
spin structure of $M^4$ can be written 
\begin{equation*}%\label{chernweil}
0\leq [c_1(L)]^2[M^4]=\frac {1}{ {4\pi ^2}}\int _{K} F_A\wedge F_A \ d\mu = \frac{1}{ {4\pi ^2}}\int _{K} (|F^+_A|^2 -| F^-_A|^2 )\ d\mu\,  .
\end{equation*}
As $F^+ _A=0$, this inequality implies that $F_A=0$, Thus the only possible solutions are flat 
connections of the $U(1)$ gauge theory. The space of flat $U(1)$ connections is determined by the 
representations of $\pi_1(M^4)$ in $U(1)$; however, these are trivial as the first betti number of $M^4$ vanishes. Consequently there are no solutions if $[c_1(L)]^2[M^4]\geq 0$.

Now suppose that $[c_1(L)]^2[M^4]$ were negative; this would imply that  $b_2^-(M^4) >0$.  Furthermore  $b_2^+(M^4)>0$ by assumption. Hence, $M^4$ has indefinite intersection form. Given this, standard results for anti-self-dual $U(1)$ connections
imply that $F^- _A= 0$ for a generic metric [see Corollary 3.21 in  \cite{FU}].  Again, it follows that $F_A = 0$ and that there are no solutions.
Therefore,  there are no nontrivial solutions of the Seiberg-Witten equations if $M^4$ admits a metric with $R> 0$. Consequently, $\sigma(M^4)\leq 0$.

The case of the asymptotically flat manifold  $(M^4-S )$ where $S$
is a finite set of points follows as in Lemma \ref{lichvanish}; by Theorem \ref{trivialunique} its compactification to $M^4$ is unique. Therefore, ${\hat {\sigma }}(M^4-S )\leq 0$. 
\end{proof}

One particularly well studied class of $4$-manifolds with nontrivial solutions to the Seiberg-Witten equations are K\"ahler $4$-manifolds, complex 2-manifolds whose K\"ahler form is closed and nondegenerate.\footnote{A hermitian metric on a complex 2-manifold can be written as $h=\sum h_{i{\bar j}}dz^i\otimes d{\bar z}^{ j}$ with complex coordinates
$\{  z^i \}$, $i=1,2$. If the 2-form $\omega =\sum h_{i{\bar j}}dz^i\wedge d{\bar z}^{ j}$
 is closed,  $d\omega =0$, and nondegenerate, then the $4$-manifold is a  K\"ahler $4$-manifold. } The following well-known result leads to a large class of $4$-manifolds with an obstruction to positive curvature \cite{Witten:1994cg}:
 
\begin{thmW}\label{SeibergWitten} Given a closed K\"ahler $4$-manifold $K$ with $b_1(K)=0$ and $b^+_2(K)\geq 2$, then the Seiberg-Witten invariant,
$SW_K\neq 0$.
\end{thmW}

Under the conditions of this theorem, the Seiberg-Witten invariant is the number of  oriented solutions of the Seiberg-Witten equations.  
The reason for assuming K\"ahler manifolds in the above theorem is that the Seiberg-Witten equations simplify on these manifolds, making the calculations easier.

Now, all K\"ahler manifolds are symplectic manifolds. It is thus natural to ask whether or not this theorem can be generalized to this case. This question was answered in the affirmative by Taubes; he extended the above theorem to symplectic 
4-manifolds \cite{Taubes}. One can think of this extension as
holding because all symplectic 4-manifolds admit an almost K\"ahler structure; they are K\"ahler manifolds up to
the requirement for transition functions to be holomorphic.

In the above theorems, the cohomology conditions are imposed to make the
monopole moduli spaces well defined. One can extend the Witten's theorem  and Taubes' theorem  to include more general cases including
$b_1(K)=0$, $b^+_2(K)=1$ and  $[c_1(L)]^2[M^4]\geq 0$ using the techniques from the proof of Theorem \ref{diffcolapse}.

Infinite families of K\"ahler manifolds with nontrivial Seiberg-Witten invariants can be constructed through blow-up. We begin by summarizing the construction of the complex surface $K'$ which is the blow-up of $K$ at $p$, following the discussion of \cite{GompfStipsicz} Ch. 2. Let $(x,y)$ be complex coordinates on ${\mathbb C}^2$ and $[u:v]$ homogeneous coordinates on ${\mathbb CP}^1$. Define the surface
$\tau =\{ ([u:v], (x,y)) \in {\mathbb CP}^1 \times {\mathbb C}^2 | xv = yu\}\subset   {\mathbb CP}^1 \times {\mathbb C}^2$. One can show that $\tau$ is diffeomorphic to  $\overline{{\mathbb CP}^2}-p$. Define the projection $\pi_2 :\tau \to {\mathbb C}^2$. The inverse image of this map a point $p$ in $C^2$ is a single point of $p\neq 0$ and ${\mathbb CP}^1$ if $p=0$.  Hence $\pi_2$ is a biholomorphism between $\tau - \pi_2^{-1}(0)$ and ${\mathbb C}^2 - \{0\}$. Next let $K$ be a complex surface and $p$ be a point in $K$. Choose a neighborhood $U$ of $p$ that is biholomorphic to an open subset $V$ of the origin in ${\mathbb C}^2$ with $p$ mapped to the origin. Then the blow-up of $K$ at $p$ is the space  $K'$  formed by removing $U$ and replacing it with $\pi_2^{-1}(V)$. 

The blow-up of a K\"ahler manifold $K$ produces another complex manifold $K'$. In addition, if $K$ is simply connected, then so is $K'$.   Consequently $K'$ is also K\"ahler as a closed complex surface is K\"ahler if and only if its first betti number is even \cite{BPV} (Also Theorem 10.1.4 in \cite{GompfStipsicz}). Furthermore, as $\tau$ is diffeomorphic to $\overline{{\mathbb CP}^2}-p$, $K'$ is diffeomorphic to the connected sum $K\# \overline{{\mathbb CP}^2}$. The connected sum does not change $b_2^+$; consequently if $K$ has nontrivial Seiberg-Witten invariant, then so does $K'$.

 K\"ahler manifolds with topology of    $k{\mathbb CP}^2\# m\overline{{\mathbb CP}^2}$ with $k\geq 2$  are of particular interest. These topological manifolds are smooth. Furthermore, there is no topological obstruction to positive scalar curvature on these manifolds. In fact, it is easy to see that they have a smooth structure that admits metrics of positive scalar curvature; take each ${\mathbb CP}^2$ and $\overline{{\mathbb CP}^2}$ factor to have the standard differentiable structure.  The Fubini-Study metric is defined with respect to this differentiable structure and has positive scalar curvature. The connected sum of two smooth manifolds can be carried out smoothly and is well known to preserve positive scalar curvature in any dimension. Therefore the standard differentiable structure on $k{\mathbb CP}^2\# m\overline{{\mathbb CP}^2}$ always admits positive scalar curvature.
 However, for sufficiently large values of $k$ and $m$, the topological manifold $k{\mathbb CP}^2\# m\overline{{\mathbb CP}^2}$ also admits other differentiable structures. Certain of these differentiable structures are induced by a corresponding  K\"ahler structure \cite{Donaldson}.
 They consequently obey the conditions of Theorem 
\ref{diffcolapse} which means that they do not admit positive scalar curvature with respect to this alternate differentiable structure.
Therefore, in these cases, the obstruction is due to the differentiable structure, not the
topology of the manifold.

Another set of examples is given by the complex surfaces $S_d$.
Let $S_d\subset {\mathbb C}P^3$ with $d>0$ an integer defined by 
$S_d=\{ (z_0,z_1,z_2,z_3)\in {\mathbb C}P^3|z_0^d+z_1^d+z_2^d+z_3^d=0 \}$. This is a smooth 
manifold which is a complex surface in ${\mathbb C}P^3$. The Lefschetz hyperplane theorem implies that $S_d$ is simply 
connected. Hence $S_d$ is K\"ahler. Using the explicit definition of $S_d$, one can find its intersection form.
When $d$ is odd, the intersection form is 
$\lambda _d \langle 1\rangle\oplus\mu _d \langle-1\rangle$
where $\lambda _d={\frac1 3}(d^3 -6d^2+11d-3)$ and $\mu _d={\frac 1 3}(d-1)(2d^2 -4d+3)$.
When $d$ is even, the intersection form is
$ l_dH\oplus m _d (-E_8) $
where $l_d={\frac 1 3}(d^3 -6d^2+11d-3)$ and $m_d={\frac 1 {24}}d(d^2 -4)$.
The second betti number is $b_2(S_d)=d(6-4d+d^2)-2$ and the signature is $\tau (S_d)={\frac1 3}(4-d^2)d$.
Since $\tau (S_d)=b_2^+(S_d)-b_2^-(S_d)$ and $b_2(S_d)=b_2^+(S_d)+b_2^-(S_d)$, it follows that 
$\tau (S_d)+b_2(S_d)=2b_2^+(S_d)$; thus a computation shows that $b_2^+(S_d)\geq 2$
%$$b_2^+(S_d)={\frac1 2}(\tau (S_d)+b_2(S_d))={\frac 1 2}({\frac 1 3}(4-d^2)d+ d(6-4d+d^2)-2)\geq 2$$
when $d\geq 4$.
Additionally, any blow-up of $S_d$ is also a complex manifold and is K\"ahler. 
Thus the surfaces $S_d$ for $d\geq 4$ and blow-ups of these surfaces are an infinite set of 
of manifolds with nontrivial Seiberg-Witten invariants.

Additional results can be derived using more powerful techniques including rational blowdown surgery  \cite{FS:1997} and knot surgery developed by Fintushel and Stern \cite{FS:1998}. These techniques were used
to construct both examples and infinite families of simply connected $4$-manifolds with exotic differentiable structures, including the case of $b^+_2 = 1$. These examples include both symplectic and smooth families.  The construction of infinitely many simply connected symplectic $4$-manifolds admitting nontrivial Seiberg-Witten invariants with $c_1^2> 8 \frac 9{10} \chi_h$ \cite{Stipsicz:1998}. More recent examples include exotic differentiable structures on $4$-manifolds with small Euler characteristic, namely ${\mathbb CP}^2 \# k \overline{{\mathbb CP}^2}$
for $k = 5,6,7,8,9$  \cite{Park:2004,SS:2004,PSS:2004} and  infinite families of
smooth $4$-manifolds homeomorphic to each other with the same Seiberg-Witten invariants \cite{FS:2004}. Further constructions of exotic differentiable structures and families of exotic differentiable structures on $3{\mathbb CP}^2\# k\overline{{\mathbb CP}^2}$ for various integers $k$
  and $(2n+2l-1){\mathbb CP}^2\# (2n+4l-1) \overline{{\mathbb CP}^2}$, $n\ge 0$, $ l \geq 1$  have also been constructed \cite{SS:2004b,ParkYun:2004,FPS:2007,ABP:2008} which have nontrivial Seiberg-Witten invariants.
  
In summary, it is clear that there are an infinite number of simply connected $4$-manifolds which have nonpositive maximal Yamabe invariant. Consequently, Theorem \ref{jangsing} implies collapse of an infinite set of topological structures  in $5$-dimensional asymptotically flat spacetimes not covered by the generalization of Gannon's theorem. 

\section{Discussion}\label{sect7}

Gannon's theorem and its generalization to higher dimensional spacetimes and Theorem \ref{jangsing} apply to different but overlapping classes of topological structures. Gannon's theorem is only applicable to asymptotically flat Cauchy surfaces with  nontrivial fundamental group; Theorem \ref{jangsing} applies to asymptotically flat Cauchy surfaces with more general topology, but that exhibit an obstruction to nonpositive curvature characterized by the maximal Yamabe invariant. Open manifolds with nontrivial fundamental group can have nonpositive maximal Yamabe invariant; punctured $n$-tori are a simple example of such manifolds in all dimensions as  the $n$-torus admits no metric with positive scalar curvature. However, manifolds such as ${\mathbb RP}^n$ admit positive scalar curvature, but have nontrivial fundamental group. Consequently punctured ${\mathbb RP}^n$ with asymptotically flat initial data collapses to form a singularity by Gannon's theorem, but is not in the class of topologies covered by Theorem \ref{jangsing}. 

In $3$ and $4$-dimensional asymptotically flat spacetime, Gannon's theorem is definitive as the fundamental group completely characterizes the  topology in these dimensions. However, this is not true in higher dimensions; Theorem \ref{jangsing} now yields singularity formation in an infinite set of simply connected asymptotically flat $5$-dimensional spacetimes, a set not covered by Gannon's theorem.  Furthermore, singularity formulation now can occur due to either the topology or the differentiable structure of the Cauchy surface. Singularity formation from differentiable structure is a novel result.

The theorem also applies in asymptotically flat spacetimes of dimension up to $8$. It may be possible to generalize Theorem \ref{jangsing} to spacetime dimension greater than $8$ by generalizing the singularity theorems to distributional apparent horizons. In higher dimensions, obstructions to positive curvature on closed simply connected manifolds is well understood. Gromov and Lawson proved that any compact simply connected manifold that does not admit a spin structure admits a metric with positive scalar curvature \cite{GL}. Hitchin proved that if a spin manifold  $M^n$ admits a metric of positive curvature then its $\alpha$- invariant vanishes, $\alpha(M^n)=0$ \cite{Hitchin}.\footnote{The  $\alpha$- invariant is  a generalization of the $\widehat A$-genus to spin manifolds.} Stolz proved that converse is also true;  in $5$ or more dimensions, any simply connected spin manifold with $\alpha(M^n)=0$ admits a metric of positive scalar curvature \cite{Stolz}. Consequently, obstructions to positive curvature are completely characterized by nonvanishing $\alpha$-invariant. 

Exotic spheres in $9$ and $10$ dimensions provide examples of manifolds with nonvanishing $\alpha$-invariant and consequently obstructions to positive scalar curvature \cite{Adams,Milnor,Hitchin}. 
However, puncturing these spaces does not yield examples of asymptotically flat manifolds with nonpositive maximal Yamabe invariant. Puncturing  an exotic $n$-sphere, $n\geq 5$,  yields ${\mathbb R}^n$ with its unique differentiable structure. Conversely,  the smooth compactification of an asymptotically flat manifold is not unique in these dimensions  by Theorem \ref{trivialunique}. Consequently, the supremum over all possible attaching maps will yield a positive maximal Yamabe invariant. However, the less robust definition of the the asymptotically flat Yamabe invariant $\hat \sigma (\Sigma^n,\Phi)$ will yield an obstruction to positive scalar curvature that depends on $\Phi$. Consequently, the obstruction now depends on the asymptotically flat initial data. 

Theorem \ref{jangsing} also differs from Gannon's theorem another respect; if $\hat \sigma(\Sigma^n)\leq 0$ then there must be  one or more apparent horizons in any  Cauchy surface $\Sigma^n$. In contrast, Gannon's theorem only implies the existence of an apparent horizon in the spacetime; it does not guarantee one in each Cauchy surface. In fact, one can exhibit Cauchy surfaces with nontrivial fundamental group and no apparent horizons.
The ${\mathbb RP}^3$ geon spacetime is a simple example. This spacetime can be constructed from the initial data for Schwarzschild spacetime on the time symmetric slice by antipodally identifying points on the minimal 2-sphere at $r=2M$. As this sphere is totally geodesic, the resulting initial data is smooth. Its evolution results in a spherically symmetric spacetime whose domain of outer communications is isomorphic to that of one asymptotic region of Schwarzschild spacetime. The ${\mathbb RP}^3$ geon has initial data with positive scalar curvature; Theorem \ref{jangsing} does not apply to this example, but as $\pi_1({\mathbb RP}^3) = {\mathbb Z}_2$, Gannon's theorem does. This spacetime clearly contains apparent horizons on Cauchy surfaces in the future of the time symmetric slice. But the time symmetric slice itself does not contain an apparent horizon; after antipodal identification the minimal 2-sphere at $r=2M$ becomes a ${\mathbb RP}^2$; a nonorientable surface. It is therefore not an apparent horizon. Clearly, Gannon's theorem cannot be improved to guarantee an apparent horizon for all topologies.

It is clearly interesting to consider whether Theorem \ref{jangsing} can be strengthened into a topological censorship theorem. The topological censorship theorems  applied to asymptotically flat spacetimes  imply that all topology associated with a nontrivial fundamental group is hidden behind horizons; i.e. that the domain of outer communications is simply connected. A corresponding strengthening of Theorem \ref{jangsing} would be that the domain of outer communications has positive maximal Yamabe invariant.

\end{document}